\documentclass[a4paper,10pt]{article}
\usepackage{jheppub}
\usepackage{orcidlink}
\usepackage{caption}
\usepackage{subcaption}
\usepackage{setspace}
\usepackage{hyperref}
\title{\boldmath Dynamics of test particles and scalar perturbations around Ay\'{o}n-Beato-Garc\'{i}a black hole coupled with cloud of strings}

\author[a,1]{Faizuddin Ahmed\orcidlink{0000-0003-2196-9622}\note{Corresponding author},}
\author[b]{Ahmad Al-Badawi\orcidlink{0000-0002-3127-3453},}
\author[c]{ \.{I}zzet Sakall{\i}\orcidlink{0000-0001-7827-9476}}
\author[d]{ and Sanjar Shaymatov\orcidlink{0000-0002-5229-7657}}

\affiliation[a]{Department of Physics, University of Science \& Technology Meghalaya, Ri-Bhoi, Meghalaya, 793101, India}
\affiliation[b]{Department of Physics, Al-Hussein Bin Talal University, 71111, Ma’an, Jordan}
\affiliation[c]{AS245 Department of Physics, Eastern Mediterranean University, 99628, Famagusta Northern Cyprus, via Mersin 10, Turkiye}
\affiliation[d]{Institute for Theoretical Physics and Cosmology, Zhejiang University of Technology, Hangzhou 310023, China\\
Institute of Fundamental and Applied Research, National Research University TIIAME, Kori Niyoziy 39, Tashkent 100000, Uzbekistan}

\emailAdd{faizuddinahmed15@gmail.com}
\emailAdd{ahmadbadawi@ahu.edu.jo}
\emailAdd{izzet.sakalli@emu.edu.tr}
\emailAdd{sanjar@astrin.uz}

\abstract{In this paper, we investigate the geodesic motion and scalar perturbations of the Ayón-Beato-García (ABG) black hole (BH) coupled with a cloud of strings (CS). By employing the effective potential approach, we analyze the trajectories of massless and massive test particles around this regular BH solution. The interplay between nonlinear electrodynamics (NLED) and CS significantly modifies the spacetime geometry, resulting in distinct dynamical properties for test particles. The behavior of null and timelike geodesics, including their stability and escape conditions, is thoroughly examined. In addition to the geodesic analysis, we study scalar perturbations by deriving the Klein-Gordon equation for massless scalar fields in this spacetime. The resulting Schrödinger-like wave equation reveals an effective potential that depends intricately on the NLED and CS parameters. Furthermore, we compute the greybody factors (GFs) to explore the energy transmission and absorption properties of theBH. Our results demonstrate that the NLED parameter $g$ and the CS parameter $\alpha$ play pivotal roles in modulating the GFs, influencing the energy spectrum of scalar radiation. These results provide significant understanding of the observational characteristics of the ABG NLED-CS BH in astrophysical contexts.       
}


\begin{document}
\maketitle
\flushbottom

\section{Introduction}\label{sec:1}

Several regular black hole (BH) solutions have been discovered through the coupling of gravity to nonlinear electrodynamics (NLED) theories. Pioneering work in this field was carried out by Born and Infeld, who introduced NLED to address the singularities associated with point charges and to mitigate energy divergence \cite{MB1}. This framework gained further significance when NLED was shown to emerge as a limiting case in specific string theory models \cite{NS1}. It is well-established that in the regime of high energies, the linearity of classical electrodynamics breaks down because of the interaction with other physical fields, making NLED a natural and promising alternative to describe electromagnetic phenomena under such extreme conditions. In these high-energy settings, NLED theories offer new perspectives by serving as a source of gravity capable of generating a wide variety of new BH solutions \cite{PP1,YM1,KE1,BP1,DVS1}. These solutions can deviate significantly from the traditional BH models, with important implications not only for BH physics but also for cosmology and string theory. In fact, NLED has found applications in both cosmological contexts \cite{VADL1,MN1,RPM1} and string theory \cite{NS1,ESF1}, reinforcing its versatility and relevance in modern theoretical physics.

The concept of regular BHs, in particular, was advanced with the work of Ayon-Beato and Garcia (ABG), who proposed the first NLED-based BH solution that satisfied the weak energy condition and asymptotically approached the Reissner-Nordstr\"{o}m BH solution \cite{EAB1}. This marked a significant step forward, since regular BHs are characterized by the absence of singularities at the center, offering a more physically plausible alternative to traditional BHs. Subsequent work further expanded this concept, with new regular exact BH solutions being derived from the coupling of NLED with Einstein's gravity \cite{EAB2, EAB5}. These solutions asymptotically mimic the Reissner-Nordstr\"{o}m BH but do not satisfy the weak energy condition, indicating the flexibility of the NLED models to accommodate different energy conditions. Further developments included the study of magnetic stable BH solutions within the context of Einstein's equations coupled to NLED \cite{EAB3}. These findings illuminate the influence of magnetic fields on the formation of stable, regular BHs. In 2005, four-parametric regular BH solutions were studied in the context of Einstein's equations coupled to NLED, providing additional information on the possible diversity of regular BH solutions \cite{EAB4}. Furthermore, NLED-sourced families of regular BH metrics have been explored in the literature \cite{LB1}, further expanding the range of viable models for regular BH in NLED theories \cite{LB2}. Some other regular BH solutions were reported in Refs. \cite{qq1,qq2,qq3,qq4,qq5,qq6,qq7}. Note that BH solutions with CS were investigated in GR as well as modified gravity theories by several authors in the literature (see, for example, Refs. \cite{AA1,JPM1,JMT1,SC1,HS1,RP1,YY1,FFN3}). 

In a curved space-time, a geodesic represents the world line of free particles influenced solely by gravity, and examining the time-like and null geodesics offers valuable insights into the geometric properties of BH space-time. In the renowned Schwarzschild space-time, particle motion can be categorized into absorption, scattering, and bound states \cite{UK}, with the bound state being particularly interesting due to its connection to the structure of celestial systems such as the solar system. Literature has explored bound geodesic motion near BHs with and without quintessence fields, as well as in the presence of string clouds \cite{FFN3, MF, OP, GM1, RFW}. Additionally, the study of the innermost stable circular orbit (ISCO) is crucial in gravitational wave astronomy, as circular orbits inside the ISCO are unstable under perturbation, often serving as initial points for the final merger of binary systems \cite{PM1, MC2}. Investigating the ISCO for a given BH reveals important information about the properties of its background geometry, since the motion of a particle is influenced by the BH's mass, charge, rotation, and any deviations due to alternative theories of gravity \cite{SSPRD1,SSPRD2}.

Cosmic strings are considered a generic outcome of symmetry-breaking phase transitions in the early universe \cite{TK1}, and they have gained significant attention due to their potential role in the formation of large-scale structures \cite{AV1, AV2}. These strings may have been present in the primordial universe and could have played a crucial role in seeding density inhomogeneities, which are fundamental to the formation of galaxies, clusters, and other cosmic structures \cite{DM1}. The theoretical idea of cosmic strings as topological defects resulting from spontaneous symmetry breaking has made them a fascinating subject of study in both cosmology and high-energy physics. In this context, Letelier \cite{PSL1} introduced the concept of CS, modeling it as a perfect fluid without pressure. This model serves as an important representation of fundamental constituents of the universe, offering profound implications from both astrophysical and cosmological perspectives. The CS model is particularly noteworthy because it has been employed as a possible material source for the Einstein field equations, leading to a generalized BH solution. This approach has been explored by several authors who have obtained exact BH solutions in the presence of string clouds. These solutions have been studied in various contexts, including general relativity \cite{PSL1, PSL2, PSL3, SHM1, DVS2, FFN1, FFN2, MC1}, as well as extended gravity theories such as Einstein-Gauss-Bonnet models \cite{EH1,JPMG1} and Lovelock gravity \cite{SGG1,NDSS,JMT1}. This body of work has significantly expanded on the pioneering contributions of Letelier \cite{PSL3}, who initially modified the Schwarzschild BH solution to accommodate a string cloud background. The study of Einstein's equations coupled with a string cloud model is particularly important because relativistic strings, when treated at the classical level, offer a robust framework for constructing models that describe fundamental cosmic structures. By including such string clouds, it is possible to better understand the role of extended objects in the dynamics of the universe, as opposed to the traditional point-particle models commonly used in theoretical physics. The universe, as a collection of such extended (non-point) objects, suggests that 1-dimensional strings, as proposed by string theory, may serve as the most plausible candidates for describing fundamental building blocks of the universe. The gravitational effects of matter in the form of clouds of cosmic and fundamental strings have thus attracted considerable attention in recent years \cite{JLS}. These investigations explore the complexities of spacetime geometry with these extended entities, aiming to comprehend their impacts on the local and global spacetime framework. Given the increasing interest in string theory and topological defects, the inclusion of such string clouds in gravitational models continues to offer valuable insights.

The computation of GFs has become an essential tool for understanding the emission spectra of BHs \cite{Kanti:2004nr,Maldacena:1996ix,MunozdeNova:2018fxv,Kanti:2002nr,Kanti:2002ge,Harmark:2007jy,Cvetic:2009jn,Boonserm:2008zg,Kanzi:2024ydp,Sekhmani:2024rpf}. GFs quantify the deviation of the Hawking radiation spectrum from that of a perfect blackbody due to the potential barrier surrounding the event horizon \cite{gf1}. These factors play a significant role in analyzing the observational signatures of BHs in high-energy astrophysical phenomena. Recent works have employed both analytical and numerical techniques to evaluate GFs for various BH solutions, particularly in the context of NLED \cite{gf2, gf3}. Inclusion of parameters such as $g$ and $\alpha$ in our study allows a detailed examination of their influence on transmission coefficients and the resulting deviations from classical blackbody radiation \cite{gf4}. Such studies \cite{Sakalli:2022xrb,Uniyal:2022xnq,Al-Badawi:2021wdm,Al-Badawi:2022aby,BarraganAmado:2024tfu,Singh:2024nvx,Heidari:2024bvd,Cui:2024uwk} are crucial for comparing theoretical predictions with observational data, potentially offering a window into the fundamental physics that governs BH dynamics \cite{gf5}. 

This investigation, therefore, seeks to study the effects of both NLED and CS on BH geometry. The interplay between these two factors may reveal new aspects of gravitational dynamics, offering insights into the formation and structure of BHs, as well as broader cosmological implications. This is the primary motivation behind the current study, which aims to contribute to our understanding of the gravitational effects induced by these fundamental cosmic objects and their role in shaping the universe. In the current study, we investigate the dynamics of massless and massive test particles around a regular charged BH in the presence of NLED, and CS has recently been reported in Ref. \cite{AK1}. The presence of NLED, as well as the CS significantly alters the behavior of the BH's gravitational field and the motion of test particles in its vicinity. In addition to the motion of test particles, we study scalar perturbation for zero spin scalar fields. Through the Klein-Gordon equation for massless scalar field, we arrive at a second-order Schrodinger-like time-independent wave equation where the effective potential depends on the aforementioned parameter NLED and CS.

The paper is organized as follows. In Sec. \ref{sec:2}, we introduce the specific ABG NLED-CS BH solution considered in this work, derived from the coupling of NLED and CS. The geodesics motion of massless and massive test particles moving in the ABG NLED-CS BH geometry are analyzed in Section \ref{sec:2.1}, with a detailed discussion of the effective potential and its implications for null and time-like geodesics. Section \ref{sec:3} investigates the scalar perturbations around the BH solution by deriving the Klein-Gordon equation and studying the scalar perturbative potential. The computation of GFs is presented in Sec. \ref{sec:3.1}, highlighting the role of $g$ and $\alpha$ in modifying the transmission coefficients. Finally, in Sec. \ref{sec:4}, we summarize our findings and discuss potential future directions for extending this research.
 
\section{ABG NLED-CS BH Space-time}\label{sec:2}

Recently, a static and spherically symmetric ABG BH in the presence of the NLED and CS was introduced in Ref. \cite{AK1}, marking an important development in BH physics. In their study, the authors explored the thermodynamics of this BH solution, finding that it obeys a modified first law of BH thermodynamics, which incorporates the effects of the NLED and CS terms. Additionally, they estimated key observable properties such as the BH's shadow and quasinormal modes, offering significant insights into its potential astrophysical signatures. In contrast, the current work aims to extend this analysis by focusing on the geodesic motion of test particles, both massless and massive, in the gravitational field of this BH. Our primary objective is to examine in detail how various factors, specifically the MM charge and the CS , influence the dynamics of these test particles, particularly their motion along geodesics. Furthermore, we explore the impact of these parameters on scalar perturbations, specifically for a zero-spin scalar field, and investigate how the MM charge and CS parameter alter the scalar perturbative potential.  

Therefore, we begin by presenting the specific static and spherically symmetric ABG NLED-CS BH solution, described by the following line element \cite{AK1}: 
\begin{equation}
    ds^2=-f(r)\,dt^2+\frac{dr^2}{f(r)}+r^2\,(d\theta^2+\sin^2 \theta),\label{bb1}
\end{equation}
where the metric function 
\begin{equation}
 f(r)=1-\alpha-\frac{2\,M\,r^2}{\left(r^2+g^2\right)^{3 / 2}}+\frac{g^2\,r^2}{\left(r^2+g^2\right)^2}.\label{bb2}
\end{equation}

The solution (\ref{bb1}) is characterized by the black hole mass MMM, the Magnetic Monopole (MM) charge $g$, and the cosmic strings (CS) parameter $\alpha$. This solution generalizes the Letelier solution \cite{EAB1, PSL1} in the absence of the MM charge ($g = 0$) and reduces to the ABG NLED-CS black hole solution when the CS parameter ($\alpha$) vanishes (with $\tilde{\alpha} = 3$ and $\beta = 4$) \cite{XCC}. Moreover, the solution (\ref{bb1}) reduces to the standard Schwarzschild black hole solution when both $\alpha = 0$ and $g = 0$.

\subsection{\large \bf Geodesic Motions of Test Particles: Null and Time-like Geodesics} \label{sec:2.1}

The dynamics of particles-whether massive or massless, charged or neutral-around a BH is one of the most fascinating and important problems in BH astrophysics. These studies provide valuable insights into the geometrical structure of spacetime and offer a deeper understanding of high-energy phenomena occurring near BHs, such as the formation of relativistic jets (which involve particles escaping the BH's gravitational influence) and accretion disks (where particles move in stable, circular orbits around the BH). The presence of strong gravitational fields, along with the influence of non-linear electrodynamics and cosmic strings, significantly alters the behavior of test particles. 

\begin{center}
    \large{\bf Effective Potential of Geodesics Motions}
\end{center}

Since the selected space-time is static and spherically symmetric, without loss of generality, we restrict ourselves to the study of equatorial geodesics motions, $\theta=\pi/2$. Therefore, the Lagrangian density function using the metric (\ref{bb1}) is given by
\begin{equation}
    -f(r)\,\dot{t}^2+\frac{1}{f(r)}\,\dot{r}^2+r^2\,\dot{\phi}^2=k,\label{cc1}
\end{equation}
where dot represents ordinary derivative w. r. t. affine parameter $\tau$ and $k=0$ for light-like geodesics and $-1$ for time-like.

One can see from space-time (\ref{bb1}) that the metric tensor $g_{\mu\nu}$ is independent of both the temporal $(t)$ and azimuthal ($\phi$) coordinates, while depends on $(r, \theta)$. Thus, there are two Killing vectors $\xi_{t} \equiv \partial_{t}$ and $\xi_{\phi}\equiv \partial_{\phi}$ associated with these coordinates ($t, \phi$), and hence, exist two conserved quantities, namely the energy ($\mathrm{E}$ ) and the angular momentum ($\mathrm{L}$). These constants of motion corresponding to these cyclic coordinates are given by
\begin{eqnarray}
    &&-\mathrm{E}=-f(r)\,\dot{t}\Rightarrow \dot{t}=\frac{\mathrm{E}}{f(r)},\label{cc2}\\
    &&\mathrm{L}=r^2\,\dot{\phi}\Rightarrow \dot{\phi}=\frac{\mathrm{L}}{r^2}.\label{cc3}
\end{eqnarray} 

With these, the geodesic equation for $r$ coordinates from Eq. (\ref{cc1}) can be obtained as,
\begin{equation}
    \dot{r}=\sqrt{E^2-V_\text{eff}(r)},\label{cc4}
\end{equation}
where the effective potential $V_\text{eff}(r)$ for null ($\epsilon=0$) or time-like geodesics ($\epsilon=-1$) is given by
\begin{equation}
    V_\text{eff} (r)=\left(-\epsilon+\frac{\mathrm{L}^2}{r^2}\right)\,\left(1-\frac{2\,M\,r^2}{\left(r^2+g^2\right)^{3 / 2}}+\frac{g^2\,r^2}{\left(r^2+g^2\right)^2}-\alpha\right).\label{cc5}
\end{equation}

In the limit where $g=0$, the metric (\ref{bb1}) reduces to the generalized Letelier solution \cite{EAB1,PSL1}. Therefore, in this limit where $g=0$, the effective potential from Eq. (\ref{cc5}) for time-like or null geodesics becomes,
\begin{equation}
    V_\text{eff} (r)=\left(-\epsilon+\frac{\mathrm{L}^2}{r^2}\right)\,\left(1-\alpha-\frac{2\,M}{r}\right).\label{cc5a}
\end{equation}

Moreover, in the limit where $\alpha=0$, without CS effects, metric (\ref{bb1}) reduces to ABG NLED-CS BH solution \cite{XCC}. Thus, in this limit where $\alpha=0$, the effective potential from Eq. (\ref{cc5}) for time-like or null geodesics reduces as,
\begin{equation}
    V_\text{eff} (r)=\left(-\epsilon+\frac{\mathrm{L}^2}{r^2}\right)\,\left(1-\frac{2\,M\,r^2}{\left(r^2+g^2\right)^{3 / 2}}+\frac{g^2\,r^2}{\left(r^2+g^2\right)^2}\right).\label{cc5b}
\end{equation}

\begin{figure}[htbp]
    \centering
    \includegraphics[width=0.4\linewidth]{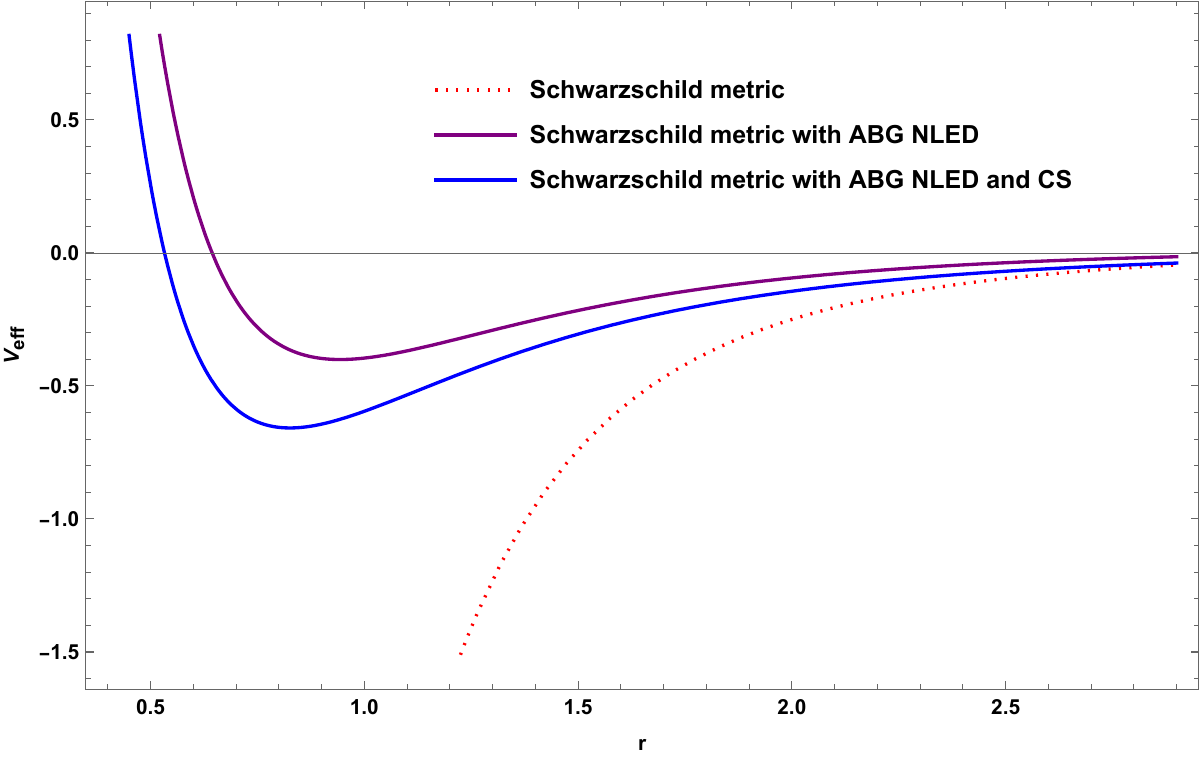}\quad\quad
    \includegraphics[width=0.4\linewidth]{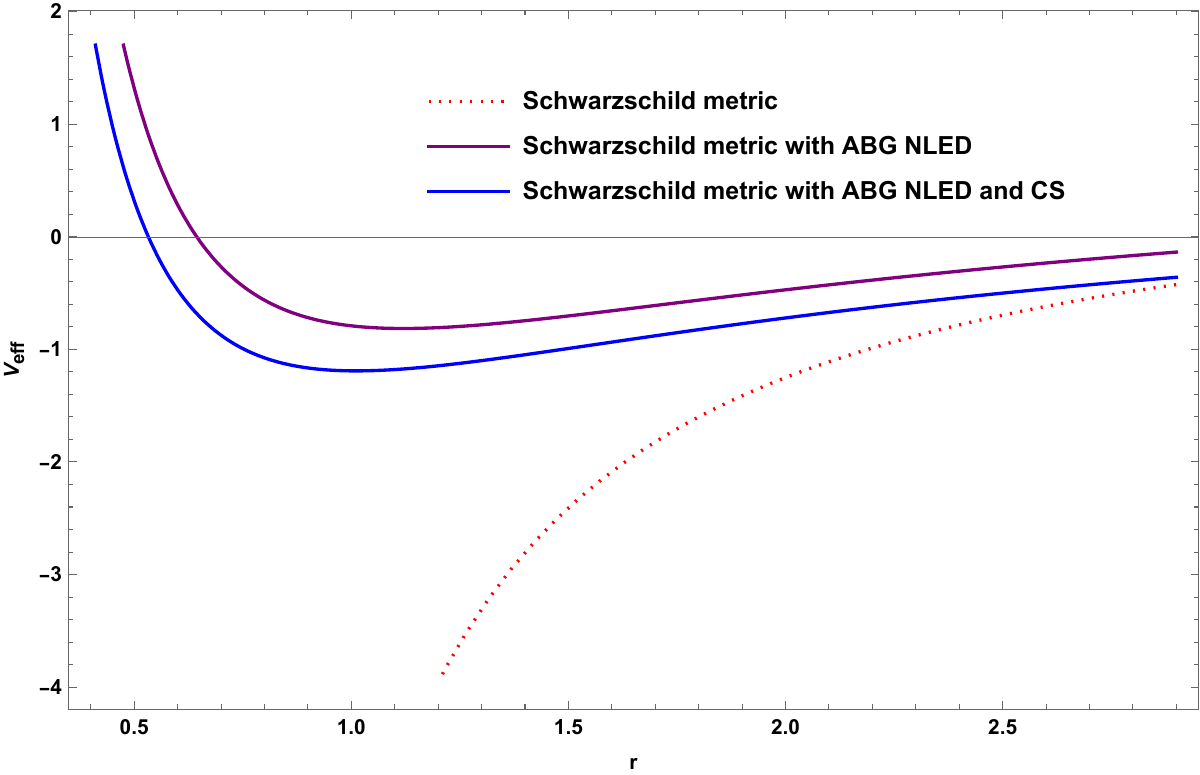}
    \caption{A comparison of the effective potential for null (left panel) and time-like (right panel) geodesics of some solutions. Here, $M=2, \mathrm{L}=1$, $\alpha=0.2$, and $g=0.9$.}
    \label{fig:3}
\end{figure}

By comparing Eqs. (\ref{cc5}), (\ref{cc5a}), and (\ref{cc5b}), it becomes clear that the MM charge $g$ and CS parameter $\alpha$ together influence the effective potential for both null and time-like geodesics. These parameters modify the effective potential in ways that distinguish the solution from that of a standard BH. In Figure (\ref{fig:3}), we present a visual comparison of the effective potential, highlighting the impact of these parameters on the dynamics of test particles in the BH's vicinity.

\begin{figure}[htbp]
    \centering
    \includegraphics[width=0.4\linewidth]{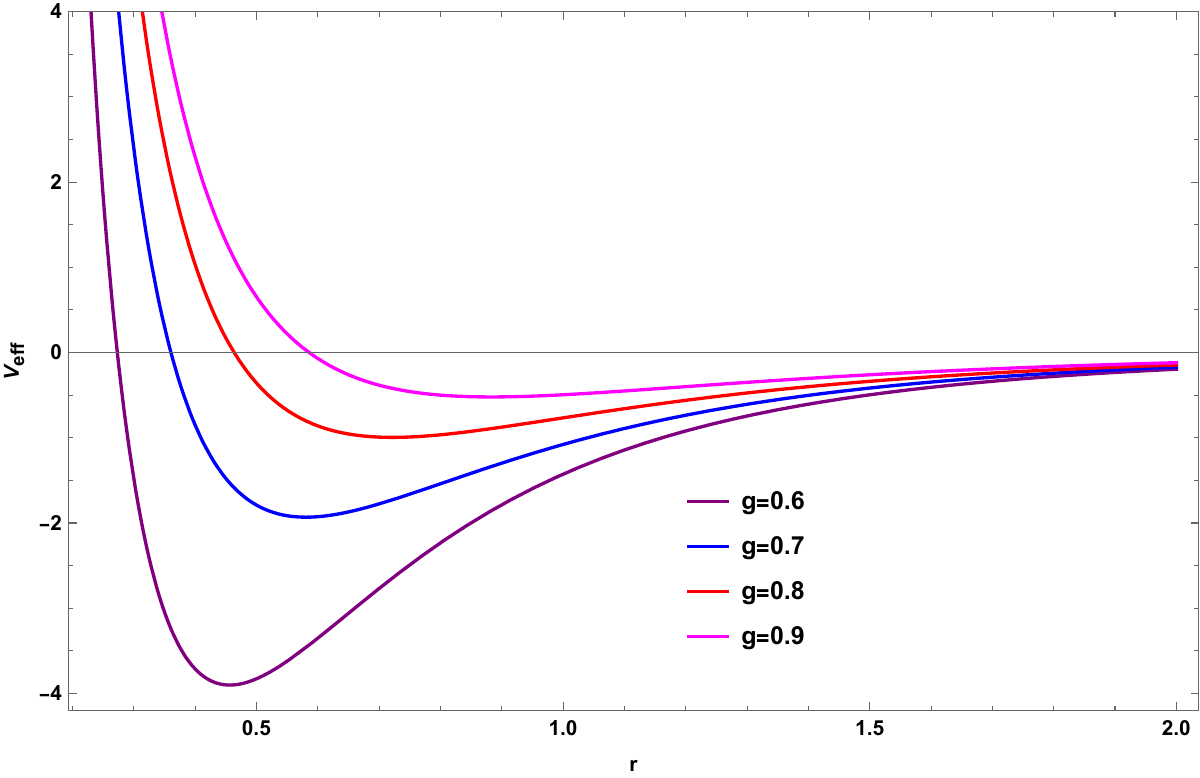}\quad\quad
    \includegraphics[width=0.4\linewidth]{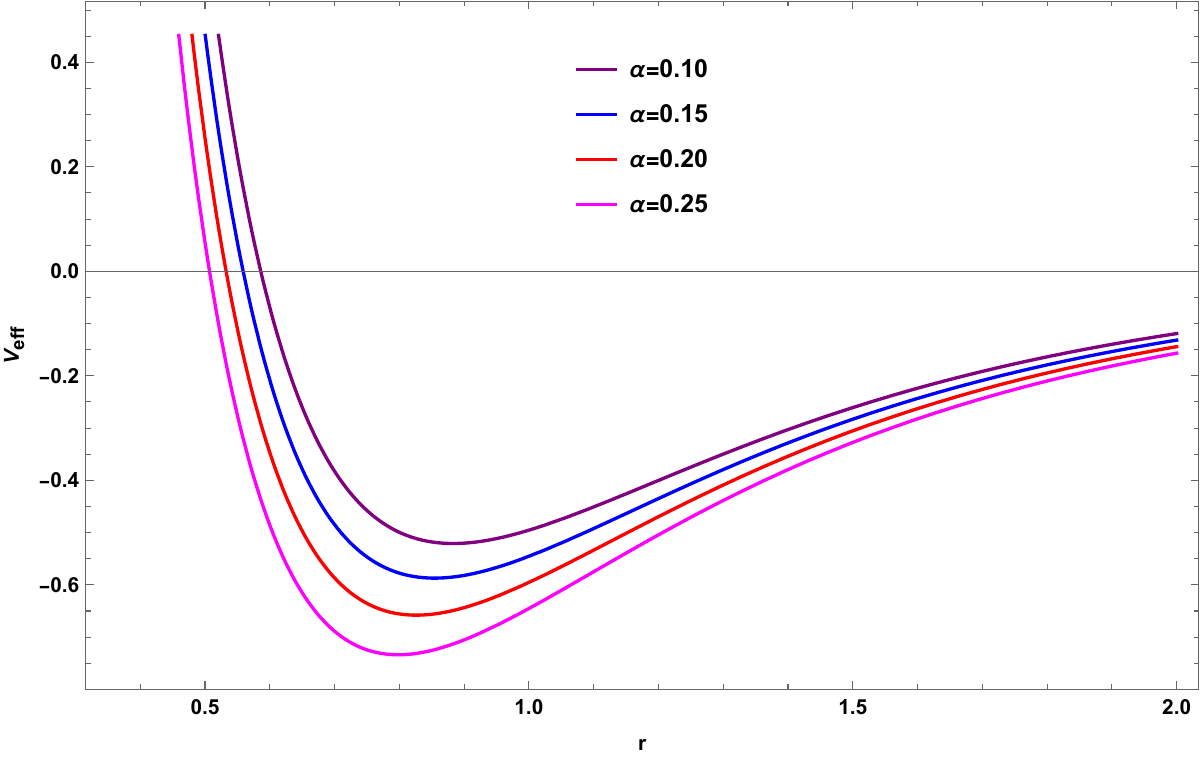}\\
    \includegraphics[width=0.4\linewidth]{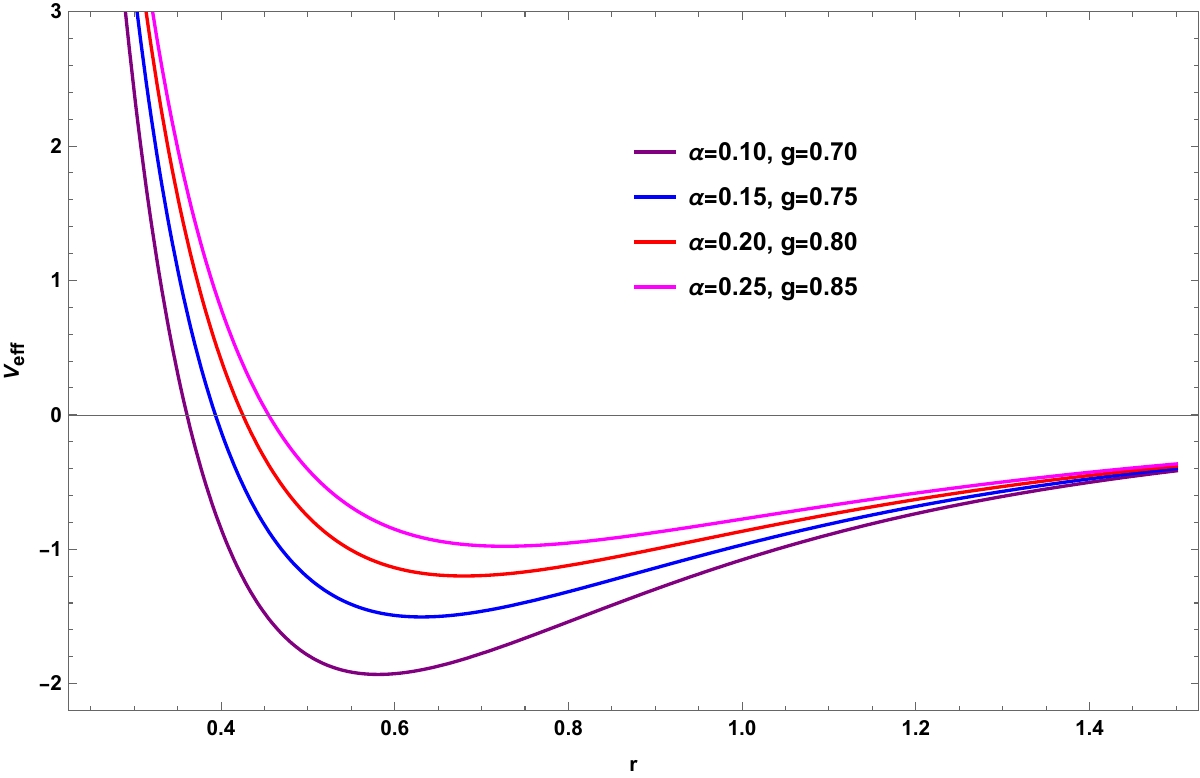}
    \caption{The behavior of the effective potential for null geodesics with different values of $\alpha$ and $g$. Here, $M=2, \mathrm{L}=1$. Top left panel: $\alpha=0.1$, Top right panel: $g=0.9$.}
    \label{fig:1}
\end{figure}

In Figure (\ref{fig:1}), we illustrate the effective potential for null geodesics as a function of the radial coordinate $r$, while varying the values of the MM charge $g$ and the CS parameter $\alpha$, with the angular momentum $\mathrm{L}=1$ kept constant. We observe that as both parameters $g$ and $\alpha$ increase, the effective potential for null geodesics also increases, indicating a modification in the dynamics of light-like particles due to the influence of these parameters on the gravitational field.

\begin{figure}[htbp]
    \centering
    \includegraphics[width=0.4\linewidth]{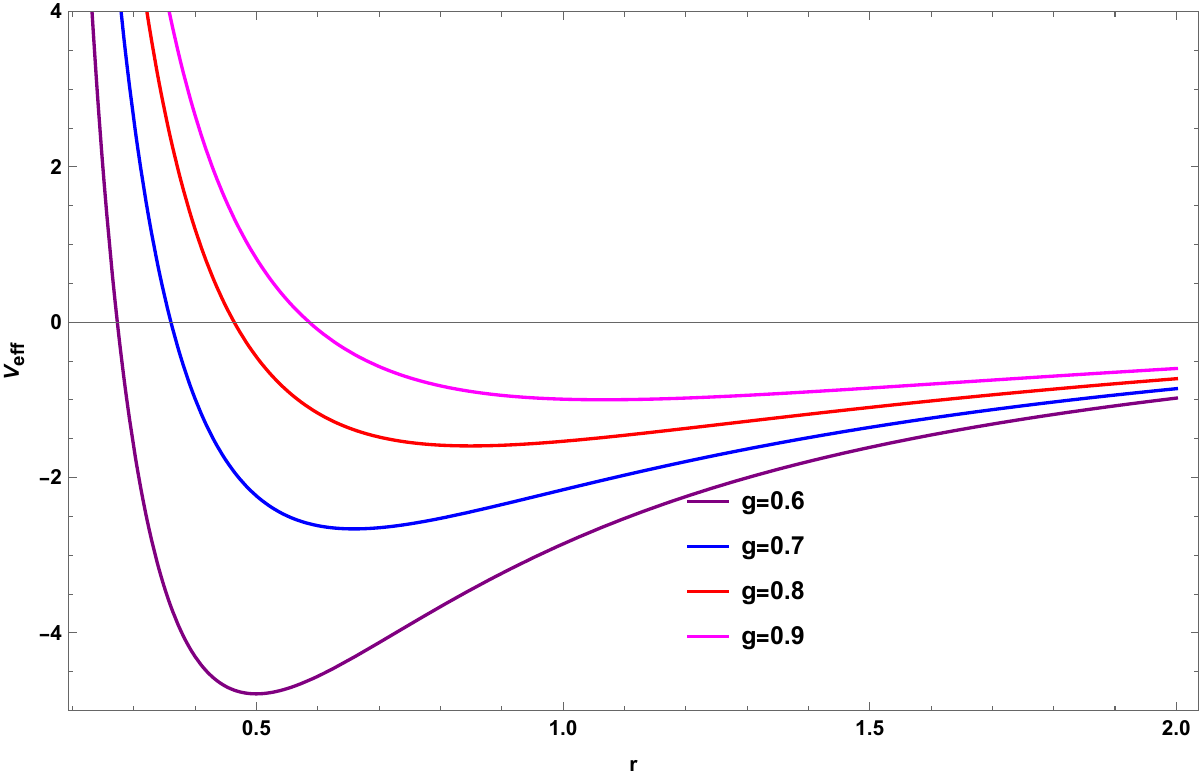}\quad\quad
    \includegraphics[width=0.4\linewidth]{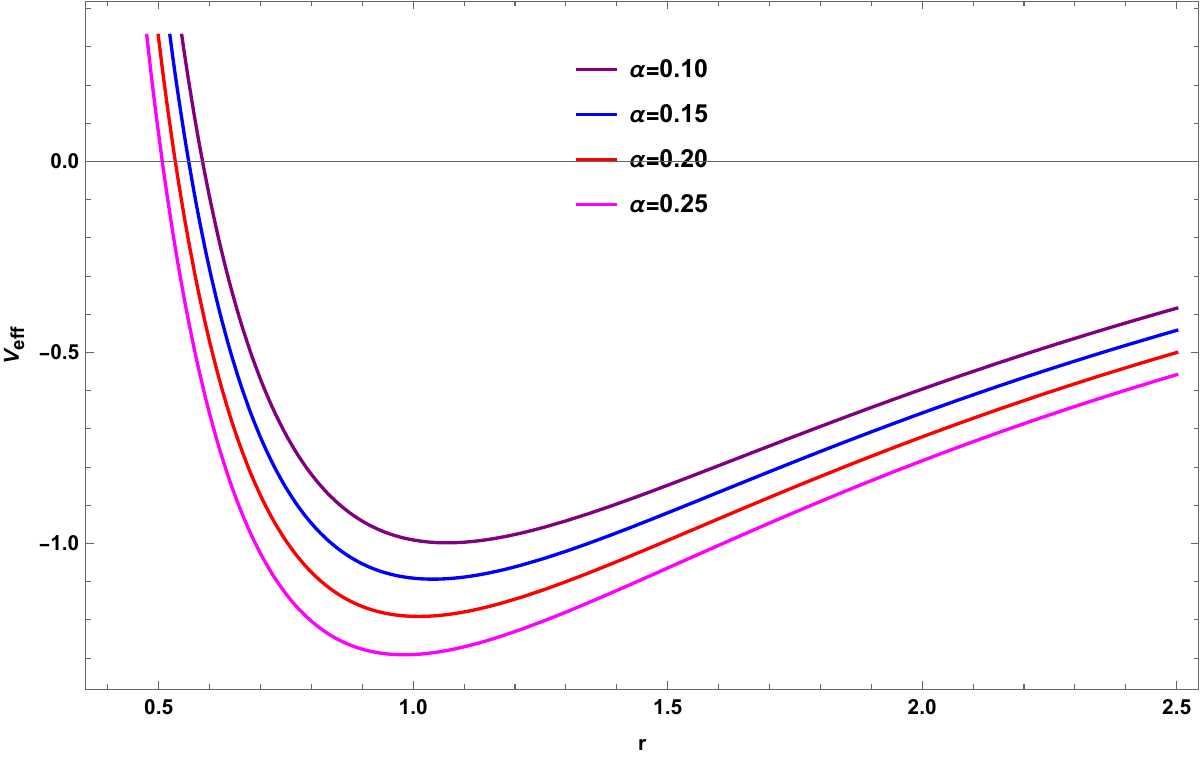}\\
    \includegraphics[width=0.4\linewidth]{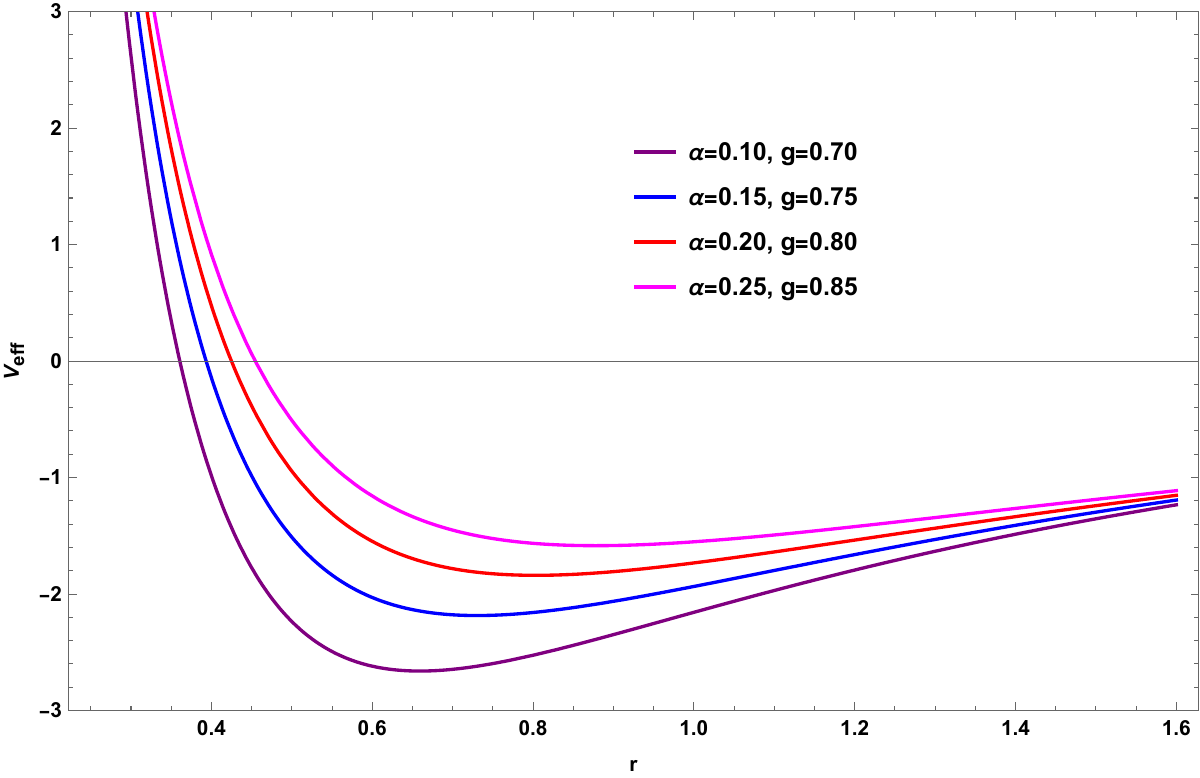}
    \caption{The behavior of the effective potential for time-like geodesics with different values of $\alpha$ and $g$. Here, $M=2, \mathrm{L}=1$. Top left panel: $\alpha=0.1$, Top right panel: $g=0.9$.}
    \label{fig:2}
\end{figure}

Similarly, in Figure (\ref{fig:2}), we generate the effective potential figure for time-like geodesics as a function of the radial coordinate $r$, while varying the values of the MM charge $g$ and the CS parameter $\alpha$, with the angular momentum $\mathrm{L}=1$ held constant. We observe that as both parameters $g$ and $\alpha$ increase, the effective potential for geodesics of time also increases.

\begin{center}
    \large{\bf Photon Light Trajectory Equation}
\end{center}

Now, we derive the photon trajectory equation under the influence of NLED and CS and see how the differential equation differs due to these parameters. Using Eqs. (\ref{cc3}) and (\ref{cc4}), we define the following quantity
\begin{equation}
    \frac{\dot{r}^2}{\dot{\phi}^2}=\left(\frac{dr}{d\phi}\right)^2=r^4\,\left[\frac{1}{\beta^2}-\frac{1}{\mathrm{L}^2}\,\left(-\epsilon+\frac{\mathrm{L}^2}{r^2}\right)\,\left(1-\frac{2\,M\,r^2}{\left(r^2+g^2\right)^{3 / 2}}+\frac{g^2\,r^2}{\left(r^2+g^2\right)^2}-\alpha\right)\right].\label{cc6}
\end{equation}

For light-like geodesics, $\epsilon=0$, from Eq. (\ref{cc6}) 
\begin{equation}
    \left(\frac{dr}{d\phi}\right)^2=r^4\,\left[\frac{1}{\beta^2}-\frac{1}{r^2}\,\left(1-\alpha-\frac{2\,M/r}{\left(1+g^2/r^2\right)^{3 / 2}}+\frac{g^2/r^2}{\left(1+g^2/r^2\right)^2}\right)\right],\label{cc7}
\end{equation}
where $\beta=\mathrm{L}/\mathrm{E}$ is the impact parameter of photon light.

Performing a transformation to new variable via $u=1/r$, Eq. (\ref{cc7}) becomes
\begin{equation}
    \left(\frac{du}{d\phi}\right)^2=\frac{1}{\beta^2}-(1-\alpha)\,u^2+\frac{2\,M\,u^3}{\left(1+u^2\,g^2\right)^{3/2}}-\frac{g^2\,u^4}{\left(1+u^2\,g^2\right)^2}.\label{cc8}
\end{equation}
In order to get a second-order differential equation form, we do a differentiating Eq. (\ref{cc8}) w. r. t. $\phi$ and after simplification yields
\begin{equation}
    \frac{d^2u}{d\phi^2}+(1-\alpha)\,u=\frac{3\,M\,u^2}{(1+u^2\,g^2)^{5/2}}-\frac{2\,g^2\,u^3}{(1+u^2\,g^2)^2}.\label{cc9}
\end{equation}
Equation (\ref{cc9}) is the photon trajectory equation in the ABG NLED-CS BH.

In the limit where $g=0$, that is, without MM charge, from Eq. (\ref{cc9}) we find
\begin{equation}
    \frac{d^2u}{d\phi^2}+(1-\alpha)\,u=3\,M\,u^2.\label{cc10}
\end{equation}
Equation (\ref{cc10}) is photon trajectory equation in BH solution with cosmic string \cite{XCC}. Noted that this equation (\ref{cc10}) is differ from the one that can be derived from the Schwarzschild spacetime with a cosmic string reported in Ref. \cite{MA}\footnote[2]{The photon light trajectory equation using the BH metric presented in Ref. \cite{MA} is given by 
$\frac{d^2u}{d\phi^2}+u=3\,\mathcal{M}\,u^2$, where $u(\phi)=\frac{1}{\alpha\,r(\phi)}$ and $\mathcal{M}=\alpha\,M$ is the physical mass of the BH \cite{DV}}.

Moreover, in the limit where $\alpha=0$, that is without CS effects, from Eq. (\ref{cc9}) we find
\begin{equation}
    \frac{d^2u}{d\phi^2}+u=\frac{3\,M\,u^2}{(1+u^2\,g^2)^{5/2}}-\frac{2\,g^2\,u^3}{(1+u^2\,g^2)^2}.\label{cc9a}
\end{equation}
Equation (\ref{cc9a}) is photon trajectory equation in generalized Letelier BH solution with cosmic string \cite{EAB1,PSL1}.

Thus, by comparing Eqs. (\ref{cc9}), (\ref{cc10}), and (\ref{cc9a}), it is evident that the MM charge $g$ and the CS parameter $\alpha$ together influence the geodesic path $r(\phi)=\frac{1}{u(\phi)}$ followed by photon ray when passing near the gravitating source. These parameters alter the photon trajectory $r(\phi)$ in a manner that distinguishes it from the path typically observed around a standard Schwarzschild BH.

\begin{center}
    \large {\bf Force on massless photon particles and circular orbits}
\end{center}

Now, we calculate the force on the massless photon particles when moving in the gravitational field produced by the chosen BH and demonstrate how the parameters, such as the MM charge $g$ and CS $\alpha$ together alter the force compared to the standard BH. The force on the massless photon particle can be defined using the effective potential $V_\text{eff}$ for light-like geodesics as follows:
\begin{equation*}
\mathrm{F}(r)=-\frac{1}{2}\frac{dV_\text{eff}\left( r\right)}{dr}.\label{cc11}
\end{equation*} 

Using the effective potential Eq. (\ref{cc5}) for null geodesics, we find the following expression of force given by
\begin{eqnarray}
    \mathrm{F}(r)=\frac{\mathrm{L}^2}{r^3}\,\left[1-\alpha-\frac{3\,M\,r^4}{(r^2+g^2)^{5/2}}+\frac{2\,g^2\,r^4}{(r^2+g^2)^3}\right].\label{cc12}
\end{eqnarray}
In the limit where $g=0$, that is without NLED effects, we find the force,
\begin{eqnarray}
    \mathrm{F}(r)=\frac{\mathrm{L}^2}{r^3}\,\left(1-\alpha-\frac{3\,M}{r}\right).\label{cc12a}
\end{eqnarray}
Moreover, in the limit where $\alpha=0$, without CS effects, this force from Eq. (\ref{cc12}) will be
\begin{eqnarray}
    \mathrm{F}(r)=\frac{\mathrm{L}^2}{r^3}\,\left[1-\frac{3\,M\,r^4}{(r^2+g^2)^{5/2}}+\frac{2\,g^2\,r^4}{(r^2+g^2)^3}\right].\label{cc12b}
\end{eqnarray}
Finally, in the limit where both parameter vanishes, $g=0=\alpha$, we find
\begin{eqnarray}
    \mathrm{F}(r)=\frac{\mathrm{L}^2}{r^3}\,\left(1-\frac{3\,M}{r}\right).\label{cc12c}
\end{eqnarray}
Equation (\ref{cc12c}) is the amount of force acting on the massless photon particles under the influence of the gravitational field produced by Schwarzschild BH solution.

Thus, from the above expressions Eqs. (\ref{cc12}), (\ref{cc12a}), (\ref{cc12b}) and (\ref{cc12c}), it clearly shows that both the MM charge parameter $g$ and the CS parameter $\alpha$ together influence the force on massless photon particles and modified the results in comparison to the standard BH solution.

\begin{figure}[htbp]
    \centering
    \includegraphics[width=0.4\linewidth]{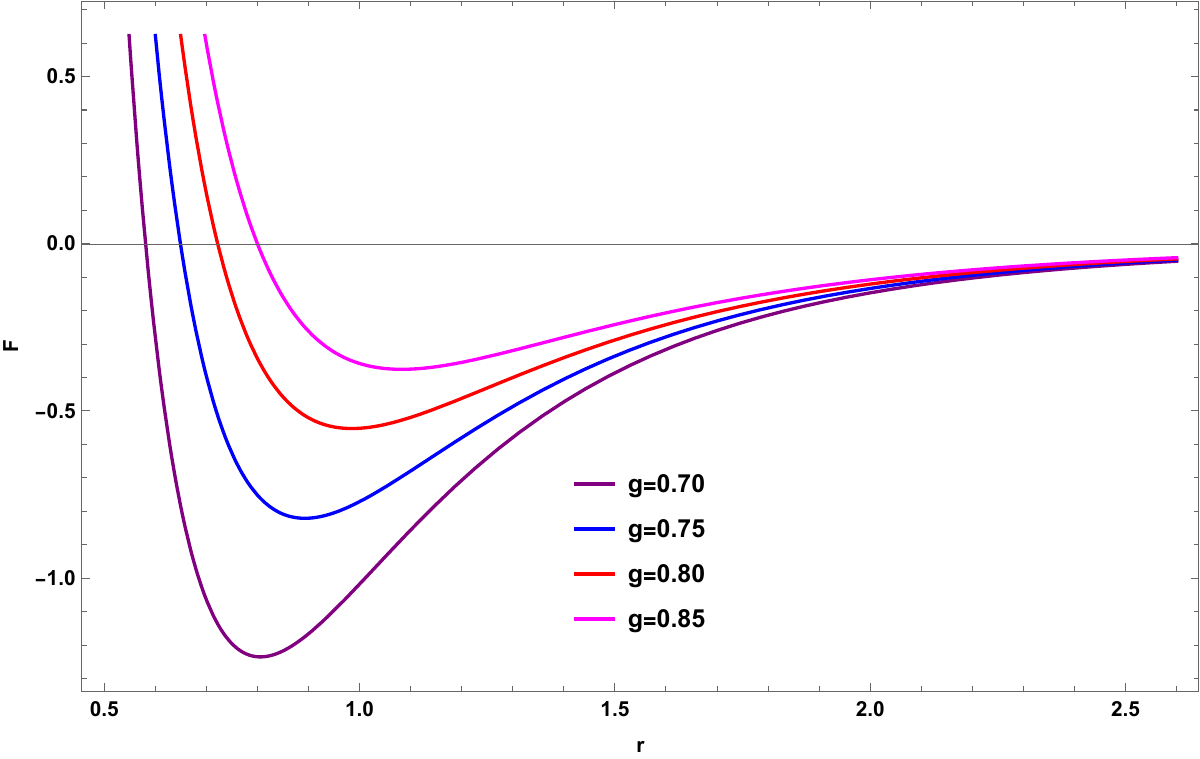}\quad\quad
    \includegraphics[width=0.4\linewidth]{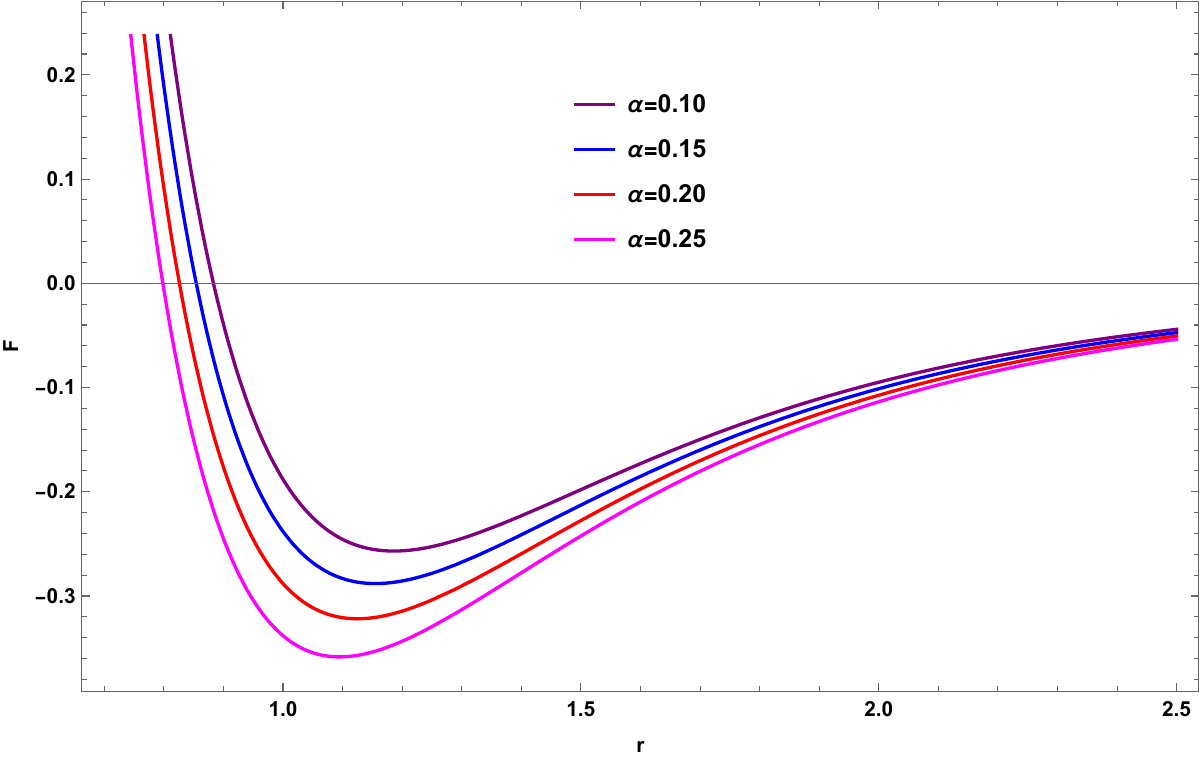}\\
    \includegraphics[width=0.4\linewidth]{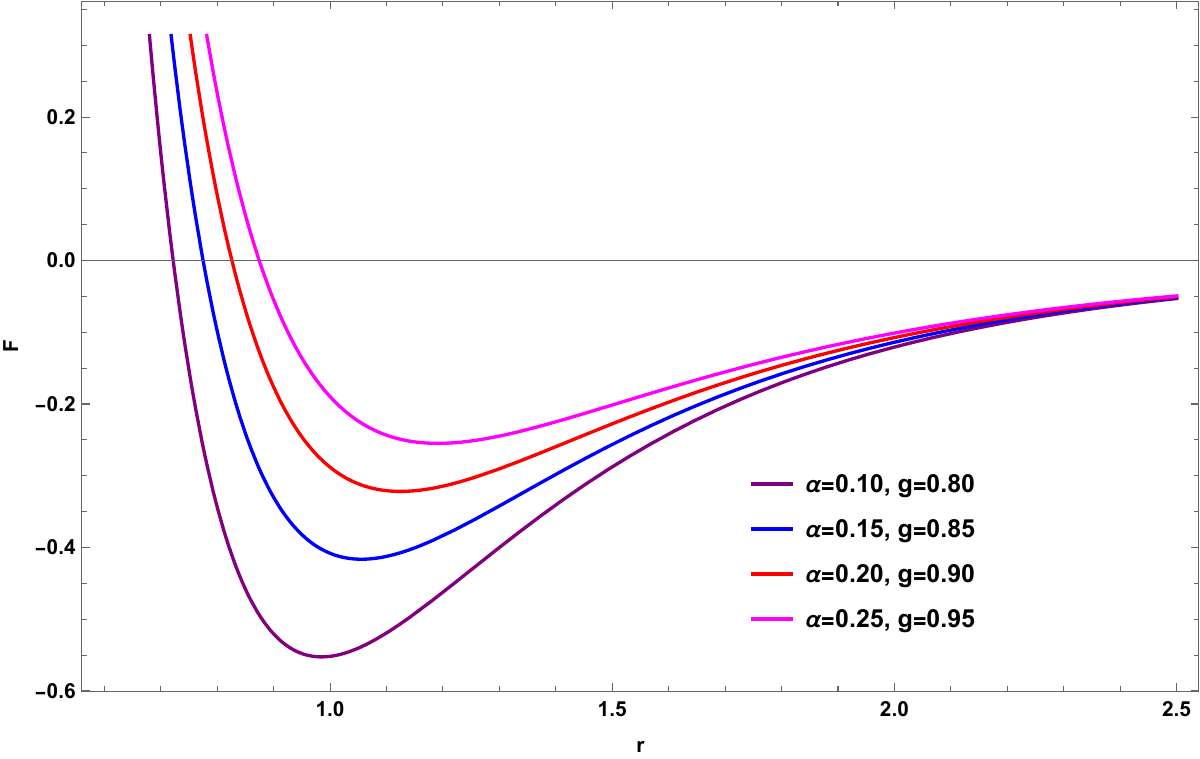}
    \caption{The behavior of force on massless photon particles as a function of $r$ with different values of $\alpha$ and $g$. Here, $M=2, \mathrm{L}=1$. Top left panel: $\alpha=0.1$, Top right panel: $g=0.9$.}
    \label{fig:4}
\end{figure}

\begin{figure}[htbp]
    \centering
    \includegraphics[width=0.45\linewidth]{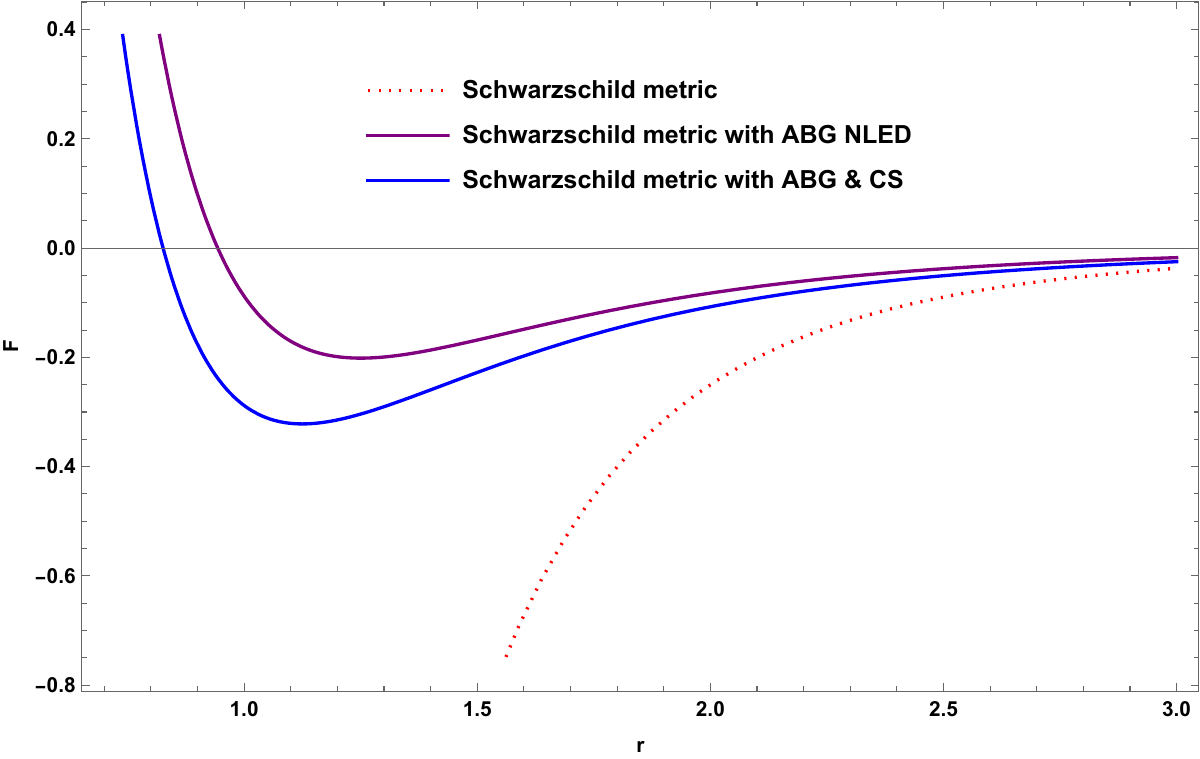}
    \caption{A comparison of force on massless photon particles as a function of $r$ for some solutions. Here, $M=2, \mathrm{L}=1$, $\alpha=0.2$, and $g=0.9$.}
    \label{fig:5}
\end{figure}

In Figure (\ref{fig:4}), we illustrate the force acting on the massless photon particles by varying the values of the MM charge parameter $g$ and the CS parameter $\alpha$. We observe that as both parameters $g$ and $\alpha$ increase, the force on the photon ray also increases. This indicates the significant influence of the gravitational field produced by the chosen BH solution, which includes the effects of non-linear electrodynamics and cosmic string contributions. In Figure (\ref{fig:5}), we compare the force acting on photon light in different BH solutions, both with and without the inclusion of NLED and CS.

For the circular photon orbit having radius $r=r_\text{ph}$, the effective potential satisfies the following conditions
\begin{equation}
    \mathrm{E}^2=V_\text{eff},\quad V'_\text{eff}(r)=0,\quad V''_\text{eff}(r)<0.\label{cc14} 
\end{equation}

Using the first condition, we find the critical impact parameter for photon light as,
\begin{equation}\label{cc15}
    \frac{\mathrm{E}^2}{\mathrm{L}^2}=\frac{1}{\beta^2_{c}}=\frac{1}{r^2_\text{ph}}\,\left(1-\alpha-\frac{2\,M\,r^2_\text{ph}}{\left(r^2_\text{ph}+g^2\right)^{3/2}}+\frac{g^2\,r^2_\text{ph}}{\left(r^2_\text{ph}+g^2\right)^2}\right).
\end{equation}

Using the second condition, the radius of the circular photon orbit can be determined by the following relation 
\begin{equation}\label{cc16}
    2\,f(r_\text{ph})=r_\text{ph}\,f'(r_\text{ph})\Rightarrow 1-\alpha-\frac{3\,M\,r^4_\text{ph}}{(r^2_\text{ph}+g^2)^{5/2}}+\frac{2\,g^2\,r^4_\text{ph}}{(r^2_\text{ph}+g^2)^3}=0.
\end{equation}

Now, using the effective potential Eq. (\ref{cc5}) for null geodesics, for the curcular null geodesics at $r=r_\text{ph}$, we find
\begin{equation}
    V''_\text{eff}(r_\text{ph})=\frac{\mathrm{L}^2}{r^4_\text{ph}}\,\left[r^2_\text{ph}\,f''(r_\text{ph})-2\,f(r_\text{ph})\right].\label{cc16a}
\end{equation}
And the coordinate angular velocity is
\begin{equation}
    \Omega_\text{ph}=\sqrt{\frac{f(r_\text{ph})}{r^2_\text{ph}}}=\sqrt{\frac{1-\alpha}{r^2_\text{ph}}-\frac{2\,M}{(r^2_\text{ph}+g^2)^{3/2}}+\frac{g^2}{(r^2_\text{ph}+g^2)^2}}.\label{cc16b}
\end{equation}
It is evident that in the limit $g=0$, the radius of circular null geodesics is $r=r_\text{ph}=\frac{3\,M}{1-\alpha}>r_\text{ph\,Sch}$. So, it is obvious that the presence of MM charge $g$ causes this radius more than that of $\frac{3\,M}{1-\alpha}$. Numerically using Eq. (\ref{cc16}), one can find radius $r_\text{ph}$ by setting different values of MM charge $g$ and CS parameter $\alpha$.

\begin{center}
    \large{\bf Dynamics of Time-like Particles}
\end{center}

Now, we focus on the dynamics of massive test particles around the selected BH. For timelike geodesics, $\epsilon=-1$ in the effective potential (\ref{cc5}). Therefore, the effective potential for timelike geodesics from Eq. (\ref{cc5}) reduces as
\begin{equation}
    V_\text{eff} (r)=\left(1+\frac{\mathrm{L}^2}{r^2}\right)\,\left(1-\frac{2\,M\,r^2}{\left(r^2+g^2\right)^{3 / 2}}+\frac{g^2\,r^2}{\left(r^2+g^2\right)^2}-\alpha\right).\label{cc17}
\end{equation}

From the condition $\dot{r}=0$ and $\ddot{r}=0$, we then determine the angular momentum and energy of test particles on the circular orbits as,
\begin{eqnarray}
    &&\mathrm{L}(r)=\sqrt{\frac{r^3\,f'(r)}{2\,f(r)-r\,f'(r)}}=\sqrt{\frac{\frac{M\,r^4\,(r^2-2\,g^2)}{(r^2+g^2)^{5/2}}-\frac{g^2\,r^4\,(r^2-g^2)}{(r^2+g^2)^3}}{\left(1-\alpha-\frac{3\,M\,r^4}{(r^2+g^2)^{5/2}}+\frac{2\,g^2\,r^4}{(r^2+g^2)^3}\right)}},\label{cc18}\\
    &&\mathrm{E}_{\pm}(r)=\pm\,\sqrt{\frac{2}{2\,f(r)-r\,f'(r)}}\,f(r)=\pm\,\frac{\left(1-\alpha-\frac{2\,M\,r^2}{\left(r^2+g^2\right)^{3 / 2}}+\frac{g^2\,r^2}{\left(r^2+g^2\right)^2}\right)}{\sqrt{1-\alpha-\frac{3\,M\,r^4}{(r^2+g^2)^{5/2}}+\frac{2\,g^2\,r^4}{(r^2+g^2)^3}}},\label{cc19}
\end{eqnarray}
which gives $\mathrm{E}=V_\text{eff}$, the effective potential. The condition $\dot{r}=0$ is termed the turning point because it gives the location at which an incoming particle turns around from the neighborhood of the gravitating source \cite{AQ1}. Equations (\ref{cc18}) and (\ref{cc19}) hold for equatorial plane only. It can be seen from (\ref{cc19}) that $\mathrm{E}_{\pm} \to \pm\,\sqrt{1-\alpha}<1$ for $r \to \infty$. Therefore, the minimum energy for the massive particle to escape from the vicinity of the BH is $\sqrt{1-\alpha}$ that depends on CS parameter.

In the limit where $g=0$, that is, without NLED effects, from Eqs. (\ref{cc18}) and (\ref{cc19}) we find the angular momentum and energy of test particles on the circular orbits as,
\begin{eqnarray}
    &&\mathrm{L}(r)=\sqrt{\frac{M\,r}{\left(1-\alpha-\frac{3\,M}{r}\right)}},\label{cc20}\\
    &&\mathrm{E}_{\pm}(r)=\pm\,\frac{\left(1-\alpha-\frac{2\,M}{r}\right)}{\sqrt{1-\alpha-\frac{3\,M}{r}}},\label{cc21}
\end{eqnarray}
Moreover, in the limit where $\alpha=0$, that is, without the cosmic string effect, from Eqs. (\ref{cc18}) and (\ref{cc19}) we find the angular momentum and energy of test particles on the circular orbits as, 
\begin{eqnarray}
    &&\mathrm{L}(r)=\sqrt{\frac{\frac{M\,r^4\,(r^2-2\,g^2)}{(r^2+g^2)^{5/2}}-\frac{g^2\,r^4\,(r^2-g^2)}{(r^2+g^2)^3}}{\left(1-\frac{3\,M\,r^4}{(r^2+g^2)^{5/2}}+\frac{2\,g^2\,r^4}{(r^2+g^2)^3}\right)}},\label{cc22}\\
    &&\mathrm{E}_{\pm}(r)=\pm\,\frac{\left(1-\frac{2\,M\,r^2}{\left(r^2+g^2\right)^{3 / 2}}+\frac{g^2\,r^2}{\left(r^2+g^2\right)^2}\right)}{\sqrt{1-\frac{3\,M\,r^4}{(r^2+g^2)^{5/2}}+\frac{2\,g^2\,r^4}{(r^2+g^2)^3}}},\label{cc23}
\end{eqnarray}

From the above expressions, Eqs. (\ref{cc18})---(\ref{cc23}), it is evident that the angular momentum and energy of the test particles on the circular orbits are influenced by both the MM charge $g$ and the CS parameter $\alpha$ together. Thus, the current result gets modified in comparison to the standard BH metric.

\begin{figure}[htbp]
    \centering
    \includegraphics[width=0.4\linewidth]{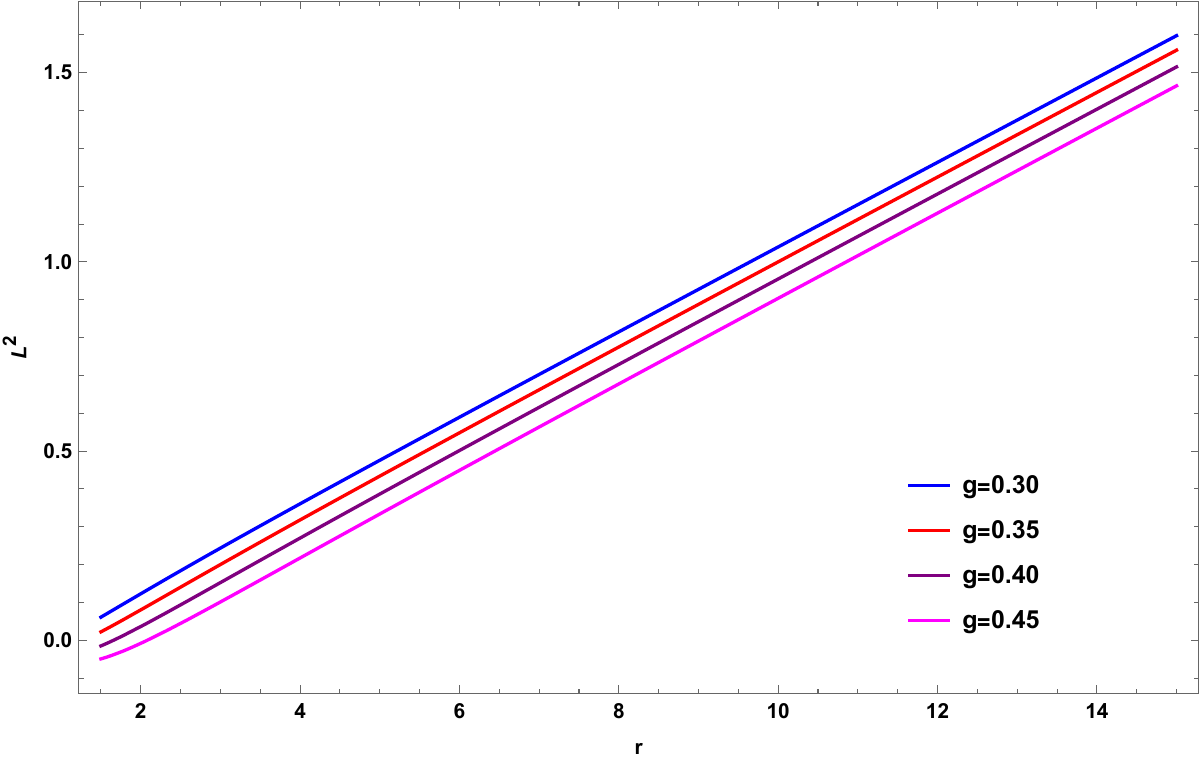}\quad\quad
    \includegraphics[width=0.4\linewidth]{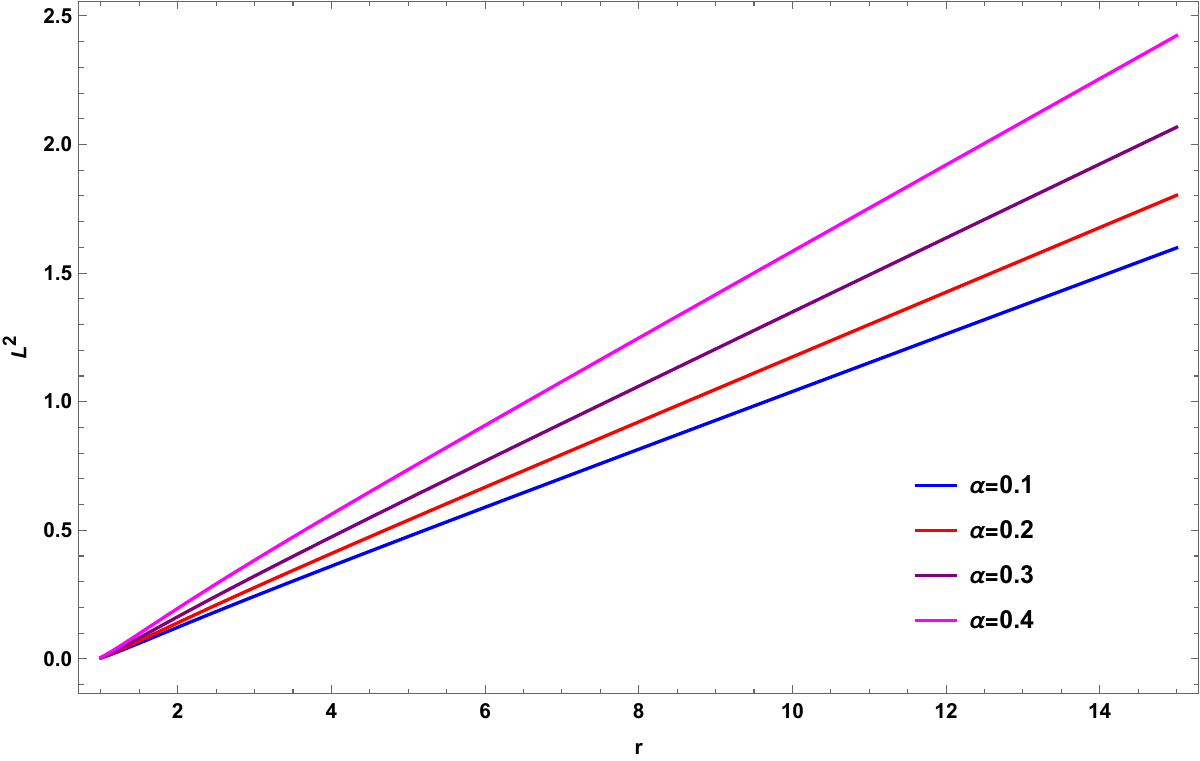}\\
    \includegraphics[width=0.4\linewidth]{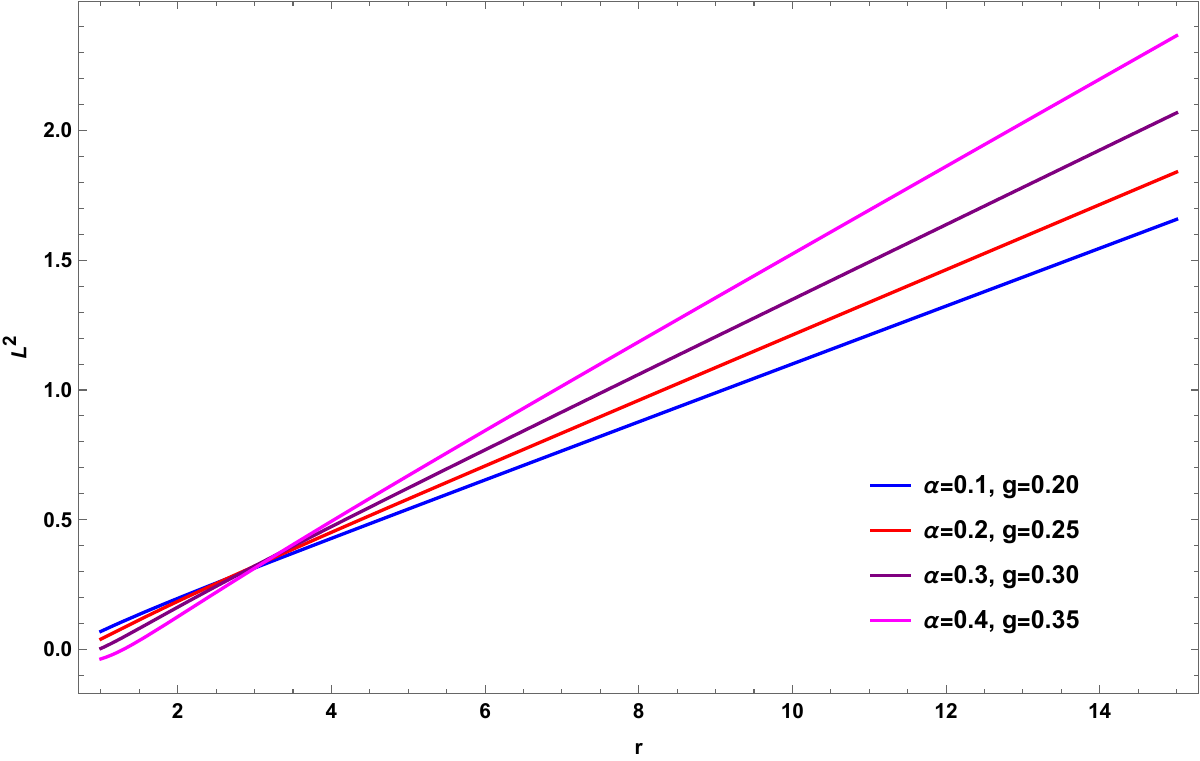}
    \caption{The behavior of $\mathrm{L}^2$ for a massive test particles as a function of $r$ with different values of $\alpha$ and $g$. Here, $M=0.1$. Top left panel: $\alpha=0.1$, Top right panel: $g=0.3$.}
    \label{fig:6}
\end{figure}

\begin{figure}[htbp]
    \centering
    \includegraphics[width=0.4\linewidth]{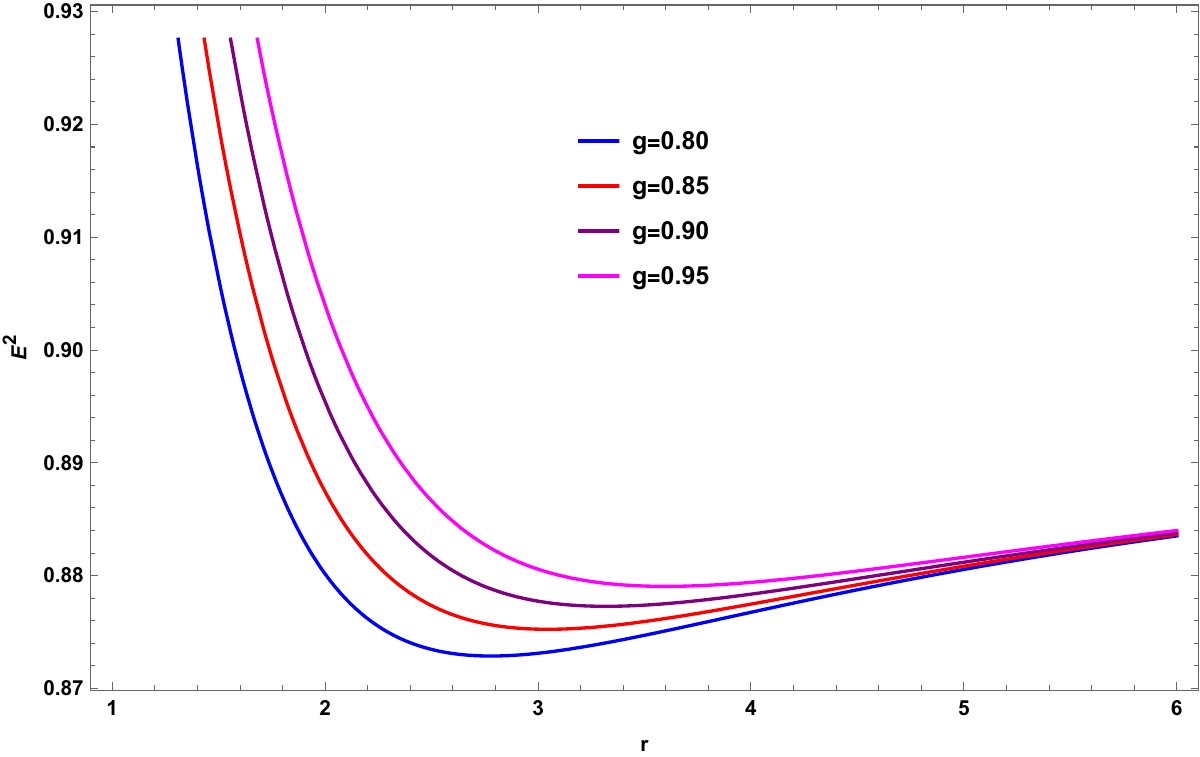}\quad\quad
    \includegraphics[width=0.4\linewidth]{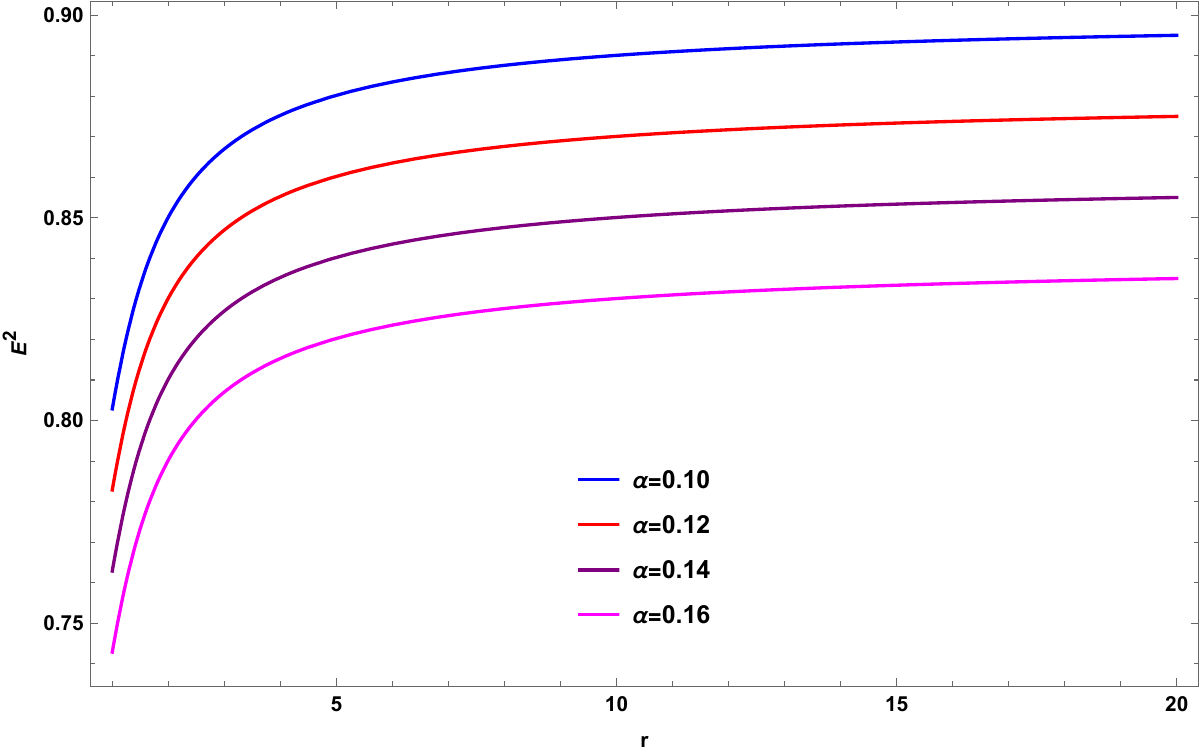}\\
    \includegraphics[width=0.4\linewidth]{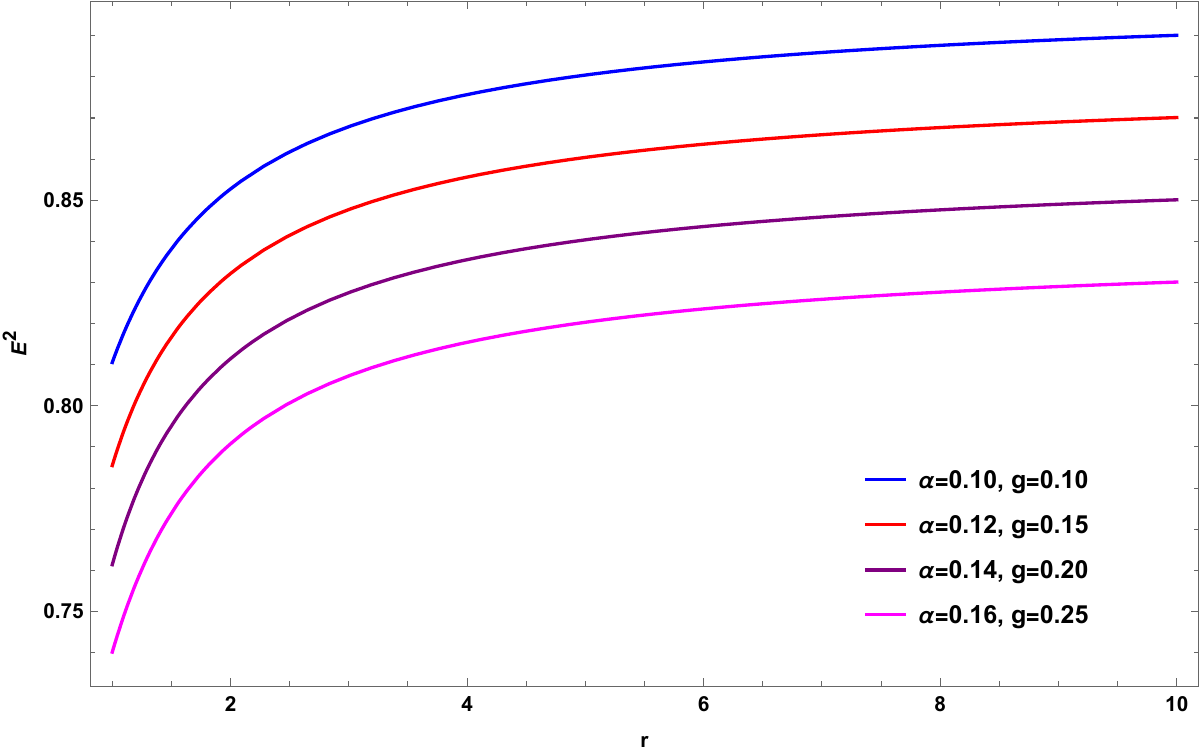}
    \caption{The behavior of $\mathrm{E}^2$ for a massive test particles as a function of $r$ with different values of $\alpha$ and $g$. Here, $M=0.1$. Top left panel: $\alpha=0.1$, Top right panel: $g=0.3$.}
    \label{fig:7}
\end{figure}

\begin{figure}[htbp]
    \centering
    \includegraphics[width=0.4\linewidth]{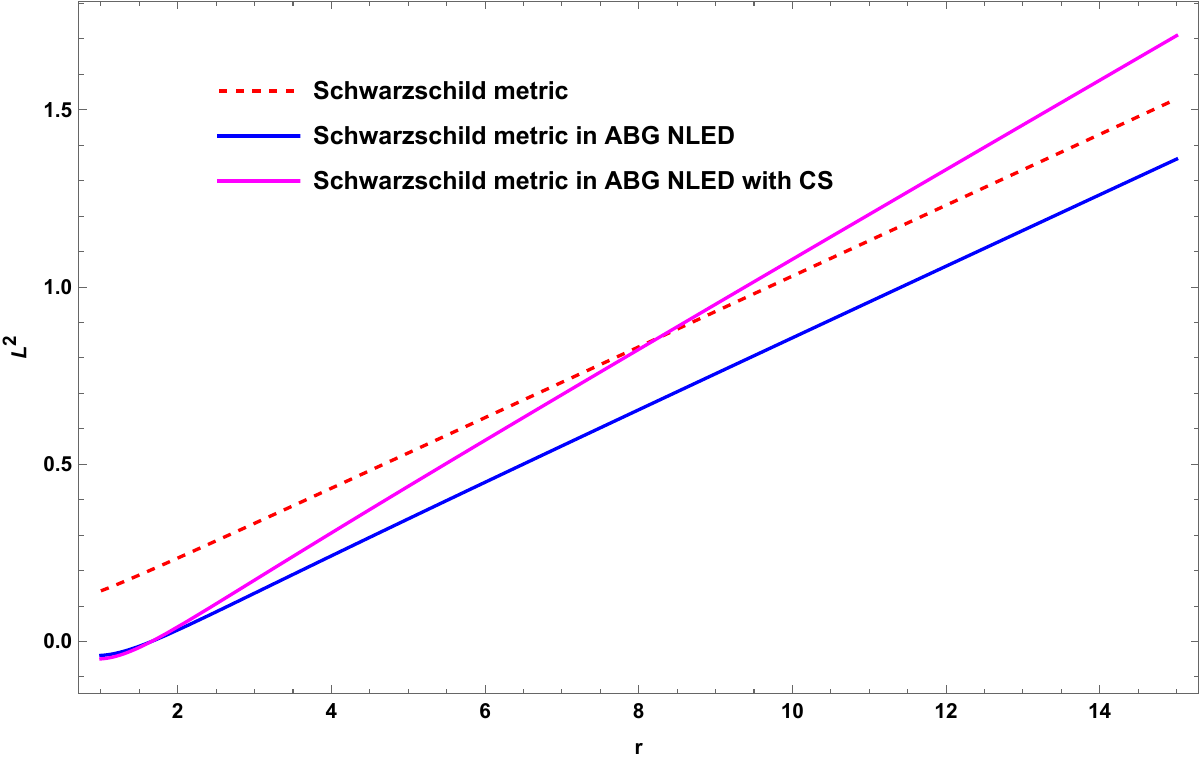}\quad\quad
    \includegraphics[width=0.4\linewidth]{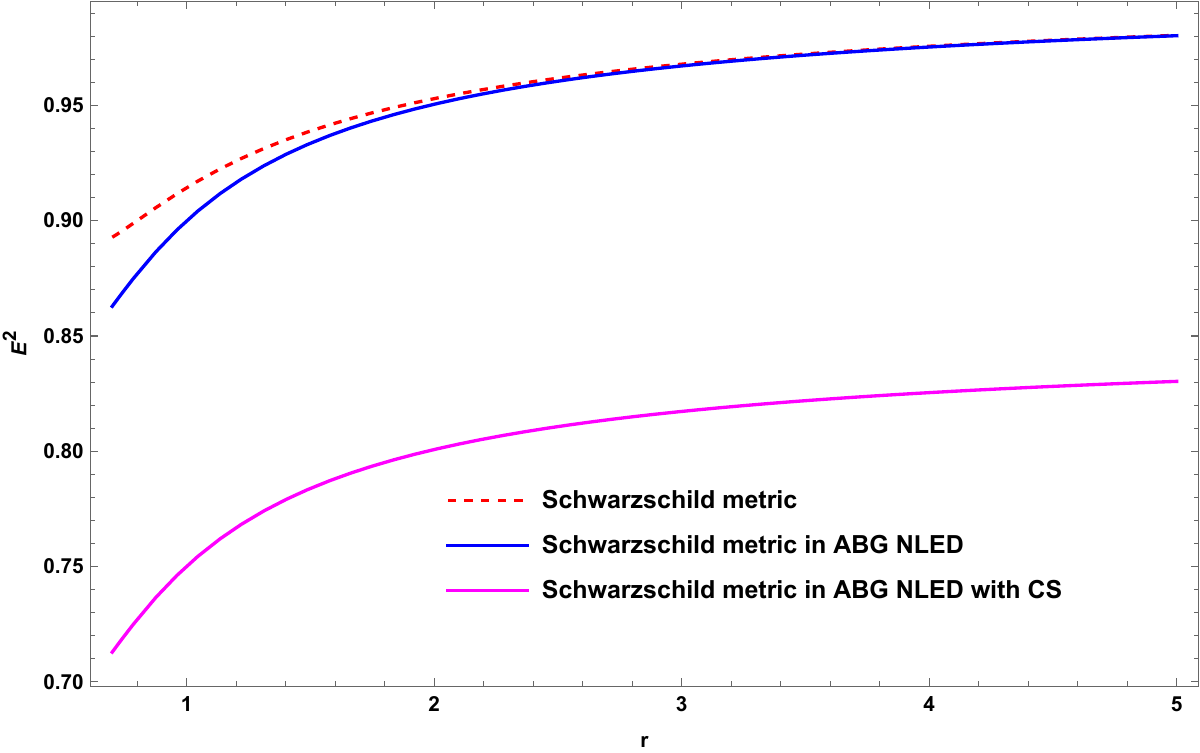}
    \caption{Comparison of $\mathrm{L}^2$ and $\mathrm{E}^2$ for massive particles as a function of $r$ for some solutions. Here, $M=0.1$. Left panel: $\alpha=0.2, g=0.4$, Right panel: $\alpha=0.15, g=0.25$.}
    \label{fig:8}
\end{figure}

In Figure (\ref{fig:6}), we illustrate the angular momentum of the test particles by varying the values of the MM charge parameter $g$ and the CS parameter $\alpha$. We observe that as both parameters $g$ and $\alpha$ increase together, the nearly linear nature of $\mathrm{L}^2$ shifts upward. Similarly, in Figure (\ref{fig:7}), we plot the energy of the test particle in the circular orbits by changing the values of the MM charge parameter $g$ and the CS parameter $\alpha$. We show that as both parameters $g$ and $\alpha$ increase together, the parabolic nature of $\mathrm{E}^2$ shifts downward. These trends indicate the influence of non-linear electrodynamics and cosmic string contributions on the angular momentum and energy of test particles in circular orbits. In Figure (\ref{fig:8}), we compare the angular momentum and energy of test particles in circular orbits for different BH solutions, both with and without the inclusion of NLED and CS.

Using Equations (\ref{cc4}) and (\ref{cc14}) and introducing new variable $u=1/r$ one can obtain the following equation
\begin{equation}
    \left(\frac{du}{d\phi}\right)^2=\frac{1}{\beta^2}-\left(\frac{1}{\mathrm{L}^2}+u^2\right)\,\left(1-\alpha-\frac{2\,M\,u}{(1+g^2\,u^2)^{3/2}}+\frac{g^2\,u^2}{(1+g^2\,u^2)^2}\right)\label{cc24}
\end{equation}
which defines the trajectory of the test particle around the ABG NLED-CS BH. Due to the difficulty in solving equation (\ref{cc24}) analytically, we solved it numerically to investigate the structure of time-like geodesics and how the MM charge $g$ and the CS parameter $\alpha$ affect the time-like geodesics in the geometry of ABG NLED-CS BH. 

\begin{figure}[htbp]
    \centering
    \includegraphics[width=0.42\linewidth]{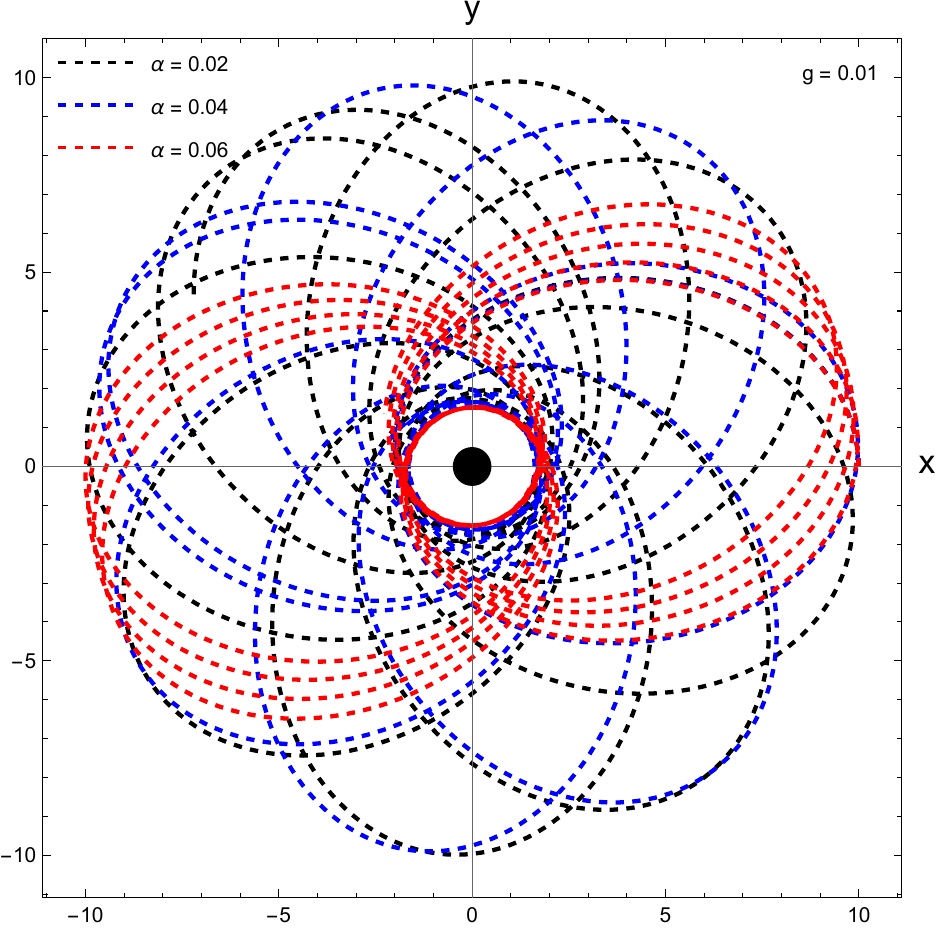}\quad\quad
    \includegraphics[width=0.42\linewidth]{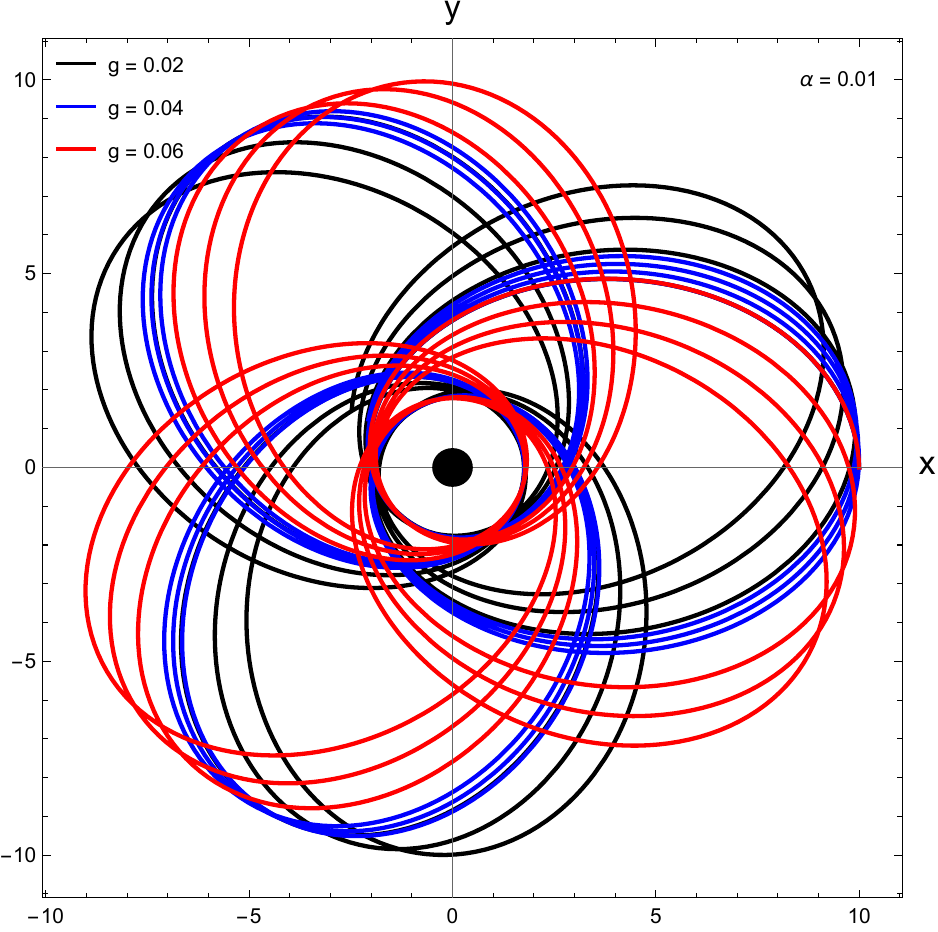}
        \caption{Time-like geodesics for a massive test particle around the  ABG NLED-CS BH, for different values of  $\alpha$ (left panel) and different values of $g$ (right panel) for the fixed energy and angular momentum.}
    \label{figa9}
\end{figure}

We plot the geodesic equation (\ref{cc24}) for a massive test particle in Figure \ref{figa9}. In Fig. \ref{figa9}, we show how a massive test particle moves around the ABG NLED-CS BH, depending on key parameters such as mass, angular momentum, energy, MM charge and the CS parameter. This analysis considers the effective potential at the maximum and minimum limiting levels. We see that the orbital plane is invariant and retains its shape, but it is caused due to the presence of the CS parameter (left panel). It is also noted that the MM charge decreases the major semimajor axis of the particle orbit whereas the CS parameter increases it. However, orbits remain stable in all scenarios as shown in both the left and right panels. 

\begin{figure*}[htbp]
\begin{center}
\begin{tabular}{cccc} 
  \includegraphics[scale=0.475]{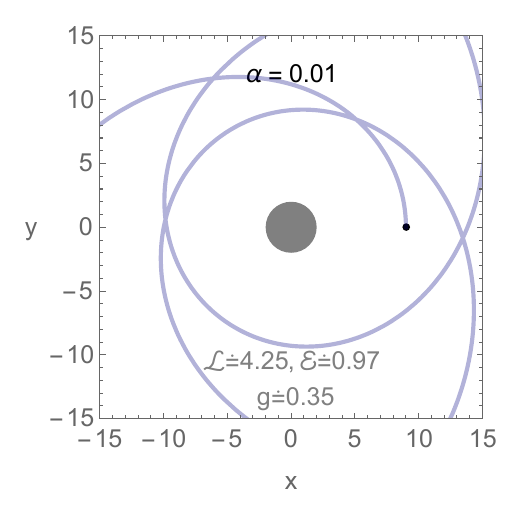}\hspace{-0.2cm}
  \includegraphics[scale=0.475]{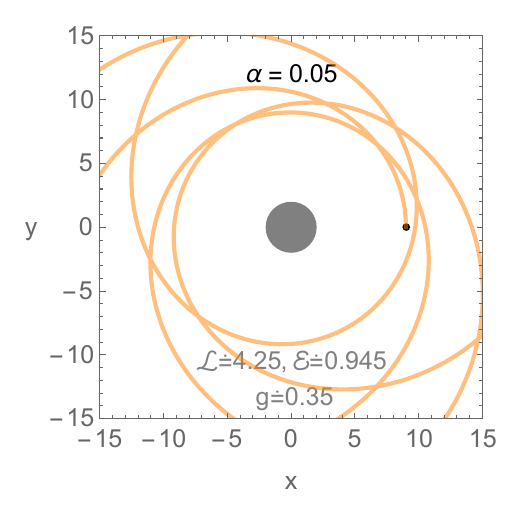}\hspace{-0.2cm}
  \includegraphics[scale=0.475]{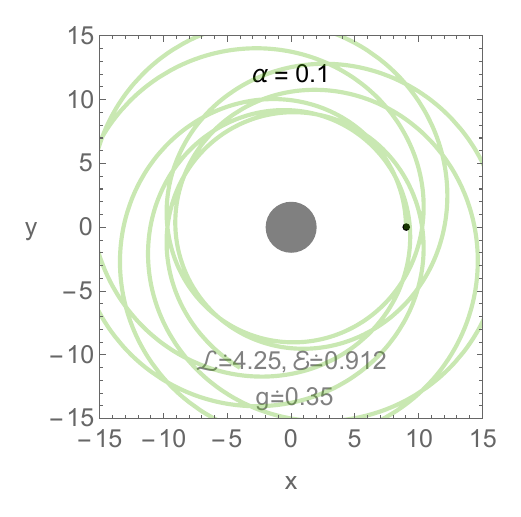}\hspace{-0.2cm}
  \includegraphics[scale=0.475]{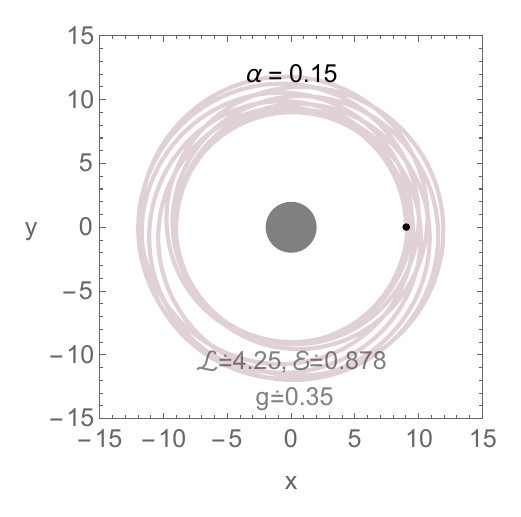}\\
  \includegraphics[scale=0.475]{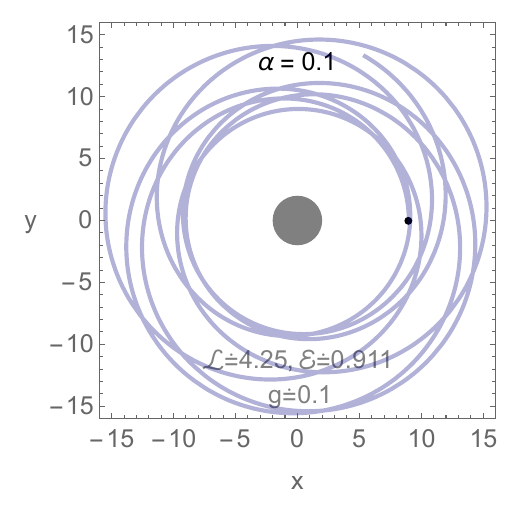}\hspace{-0.2cm}
  \includegraphics[scale=0.475]{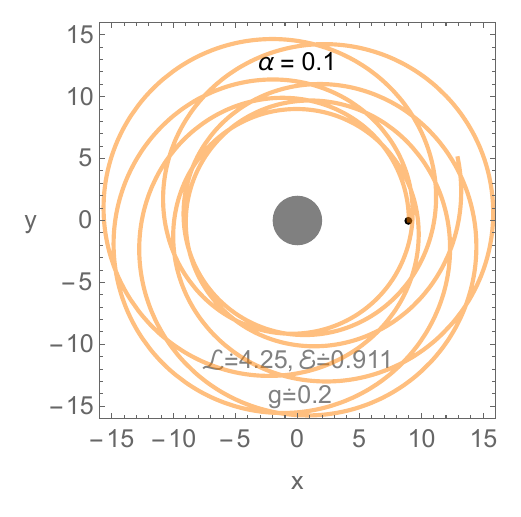}\hspace{-0.2cm}
  \includegraphics[scale=0.475]{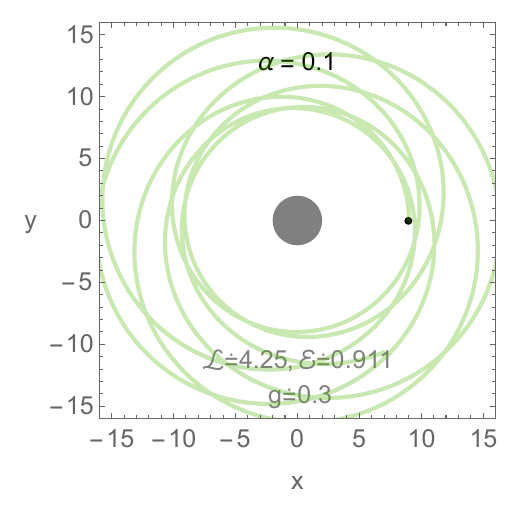}\hspace{-0.2cm}
  \includegraphics[scale=0.475]{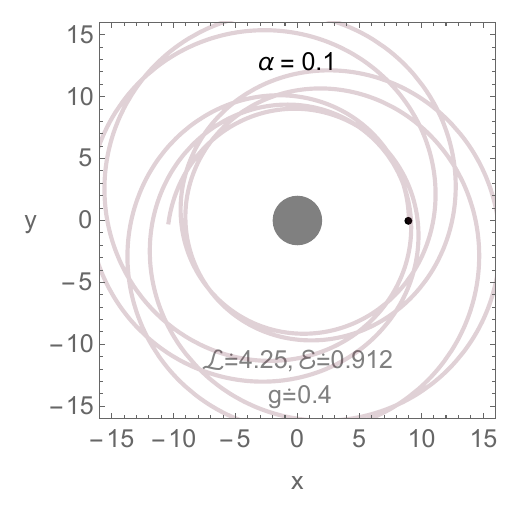}
  \end{tabular}
\caption{\label{fig:tm} 
Time-like geodesics for massive test particles at various energies around the ABG NLED-CS BH, for different values of $\alpha$ (top panels) and different values of $g$ (bottom panels). Here, we note that the behavior of time-like geodesics for massive particles are observed from the polar view ($z = 0$). } 
\end{center}
\end{figure*}

To understand the orbital behavior of massive particles around the ABG NLED-CS BH, we investigate the influence of BH parameters on time-like geodesics for massive particles at various energies. Focusing on equatorial plane motion around the ABG NLED-CS BH for simplicity, we qualitatively analyze the time-like geodesics, illustrating their behavior in Fig.~\ref{fig:tm}. As shown, the observed orbits-escaping and  bounded depending on the BH parameters, providing insights into their impact near the BH. The top row of Fig.~\ref{fig:tm} demonstrates a transition from initially escaping orbits to bounded orbits with a slight increase in $\alpha$ (keeping $g$ constant). Similarly, the bottom row illustrates a shift from bounded to escaping orbits as $g$ increases from 0.1 to 0.4 in equal intervals (keeping $\alpha$ constant). 

A critical circular orbit that is particularly interesting is the innermost stable circular orbit (ISCO) which is determined by equations in (\ref{cc14}) and an additional equation $V_\text{eff}'' = 0$ where $V_\text{eff}$ is given by Eq. (\ref{cc17}).  Based on the three conditions above, the ISCO radius equation can be derived from the following equation: 
\begin{equation}
\frac{4\,f\,f'\,}{r}-\frac{f\,f'\,}{r}+f\,f''-2\,f'^2=0.
\label{isco} 
\end{equation}
The above equation (\ref{isco}) is extremely difficult to solve analytically and thus can be solved numerically. The numerical values of ISCO are shown in Table \ref{taba1}
for different values of MM charge $(g)$ and CS parameter $(\alpha)$.
\begin{center}
\begin{tabular}{|c|c|c|c|c|c|}
 \hline 
 \multicolumn{6}{|c|}{ $r_\text{ISCO}$}
\\ \hline 
$\alpha$& $g =0$ & $0.1$ & $0.3$ & $0.5$ & $0.7$\\ \hline
$0.0$ & $6.00000$ & $5.96904$ & $5.71147$ & $5.12644$ & $3.83195$ \\ 
$0.1$ & $6.66667$ & $6.63732$ & $6.39471$ & $5.85723$ & $4.81129$ \\ 
$0.3$ & $8.57143$ & $8.54528$ & $8.33155$ & $7.87583$ & $7.09391$ \\ 
$0.5$ & $12.0000$ & $11.9771$ & $11.7910$ & $11.4051$ & $10.7844$ \\ 
 \hline
\end{tabular}
\captionof{table}{For different values of parameters $\alpha$ and $g$ with $M=1$, the innermost stable circular orbit has been tabulated numerically for  ABG NLED-CS BH.} \label{taba1}
\end{center}

\begin{figure}
    \centering
    \includegraphics[width=0.5\linewidth]{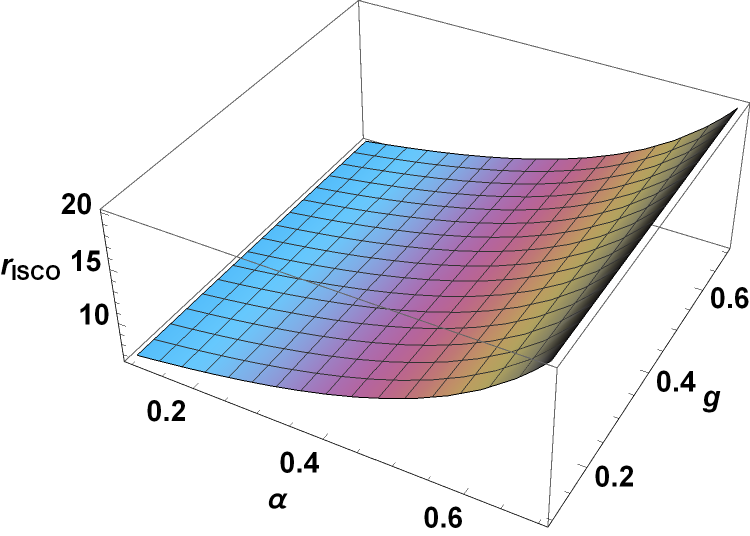}
    \caption{The plot of the ISCO versus the MM charge $g$ and the CS parameter $\alpha$ with $M=1$.}
    \label{figa1}
\end{figure}

Table \ref{taba1} shows that for a given value of the MM charge $g$, as the CS parameter $\alpha$ increases, the ISCO radius increases significantly.  For a fixed value $\alpha$, the radius of the ISCO decreases as $g$ increases. Therefore, both parameters have the opposite effect on ISCO. We showed that the effects of the MM charge and the CS parameter are opposing (see Fig. \ref{figa1}).

Finally, we now calculate the orbital angular velocity of time-like particles on the circular orbits of radius $r=r_0$, which is given by \cite{SSPRD1,SSPS}
\begin{equation}
    \Omega=\sqrt{\frac{f'(r)}{2\,r}}\Bigg{|}_{r=r_0}.\label{vel1}
\end{equation}
In our case, we find this velocity will be
\begin{equation}
    \Omega=\sqrt{\frac{1}{(r^2_0+g^2)^3}\,\left[M\,\left(r^2_0-2\,g^2\right)\,\sqrt{r^2_0+g^2}-g^2\,\left(r^2_0-g^2\right)\right]}.\label{vel2}
\end{equation}
In the limit where $g=0$, this velocity reduces to the result, $\Omega=\sqrt{M/r^3_0}$. From the above expression, Eq. (\ref{vel2}), it is evident that the angular velocity of time-like particles on circular orbits is influenced only by the MM charge $g$.

\begin{center}
    \large{\bf Stability of time-like particle orbits}
\end{center}

Now, we focus on stability analysis of the time-like geodesics. The stability of the circular orbits for massive particles can be studied using the Lyapunov exponent, on average, represents the numerical characteristics of the exponential divergence rate of neighboring orbits in phase space \cite{DL}. The positive Lyapunov exponent indicates the divergence between neighboring orbits with the emergence of chaos in the system. In comparison, the negative values indicate the convergence of the orbits. The expression of the Lyapunov exponent in terms of the effective potential is \cite{VC,JR}
\begin{equation}
    \lambda_{\mathrm{L}}=\frac{1}{\dot{t}}\,\sqrt{-\frac{1}{2}\,\partial^2_{r}\,V_\text{eff}},\label{kk1}
\end{equation}
where $V_\text{eff}$ is the effective potential for time-like geodesics given in Eq. (\ref{cc17}) and
\begin{equation}
    \dot{t}=\frac{\mathrm{E}}{f(r)}=\sqrt{\frac{2}{2\,f(r)-r\,f'(r)}}=\frac{1}{\sqrt{1-\alpha-\frac{3\,M\,r^4}{(r^2+g^2)^{5/2}}+\frac{2\,g^2\,r^4}{(r^2+g^2)^3}}}.\label{kk2}
\end{equation}

The stability of the particle orbits can be determined by the positive and negative values of the Lyapunov exponent $\lambda_{\mathrm{L}}$. More specifically, the orbit remains stable for $\lambda_{\mathrm{L}}<0$, while it becomes chaotic for $\lambda_{\mathrm{L}}>0$.

Using Eqs. (\ref{kk2}) and (\ref{cc17}) into the Eq. (\ref{kk1}) and after simplification, we get the following expression in terms of the metric function as \cite{VC}, 
\begin{equation}
    \lambda^2_{\mathrm{L}}=\frac{1}{2}\,\left[2\,(f'(r))^2-f(r)\,f''(r)-\frac{3\,f(r)\,f'(r)}{r}\right].\label{kk3}
\end{equation}

Thereby, using the metric function (\ref{bb2}) into the expression (\ref{kk3}) and after simplification , we find the final result as follows: 
{\small
\begin{eqnarray}
    &&\lambda_{\mathrm{L}}=\frac{1}{g^2 + r^2)^3}\,\Bigg[M\,r^8\, \left(6\, M +\sqrt{g^2 + r^2}\, (-1 +\alpha)\right)+ 
  g^4\, r^4\, \left(r^2\, (12 - 8\, \alpha) + 
     9\, M\, \sqrt{g^2 + r^2}\,(-1 +\alpha)\right)\nonumber\\
     &&+4\,g^{10}\, (-1 +\alpha)- 
  8\, g^8\, M\, \sqrt{g^2 + r^2}\,(-1 +\alpha)- 
  4\, g^6\, r^2\, \left(3\, r^2 + 2\, M\, \sqrt{g^2 + r^2}\right)\,(-1 + \alpha)\nonumber\\
  &&+g^2\, M\, r^6\, \left(6\, M + \sqrt{g^2 + r^2}(-19 + 10\, \alpha)\right)\Bigg]^{1/2}.
\end{eqnarray}
\normalsize}

\begin{figure}[htbp]
    \centering
    \includegraphics[width=0.45\linewidth]{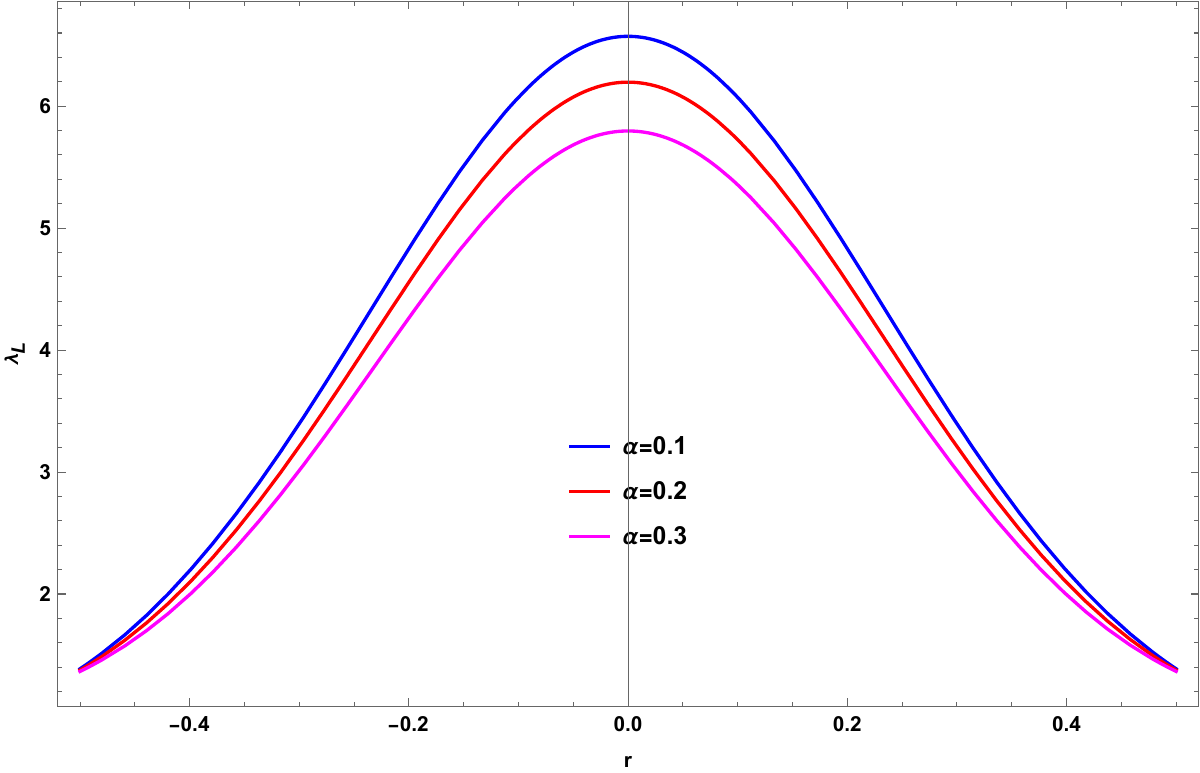}\quad\quad
    \includegraphics[width=0.45\linewidth]{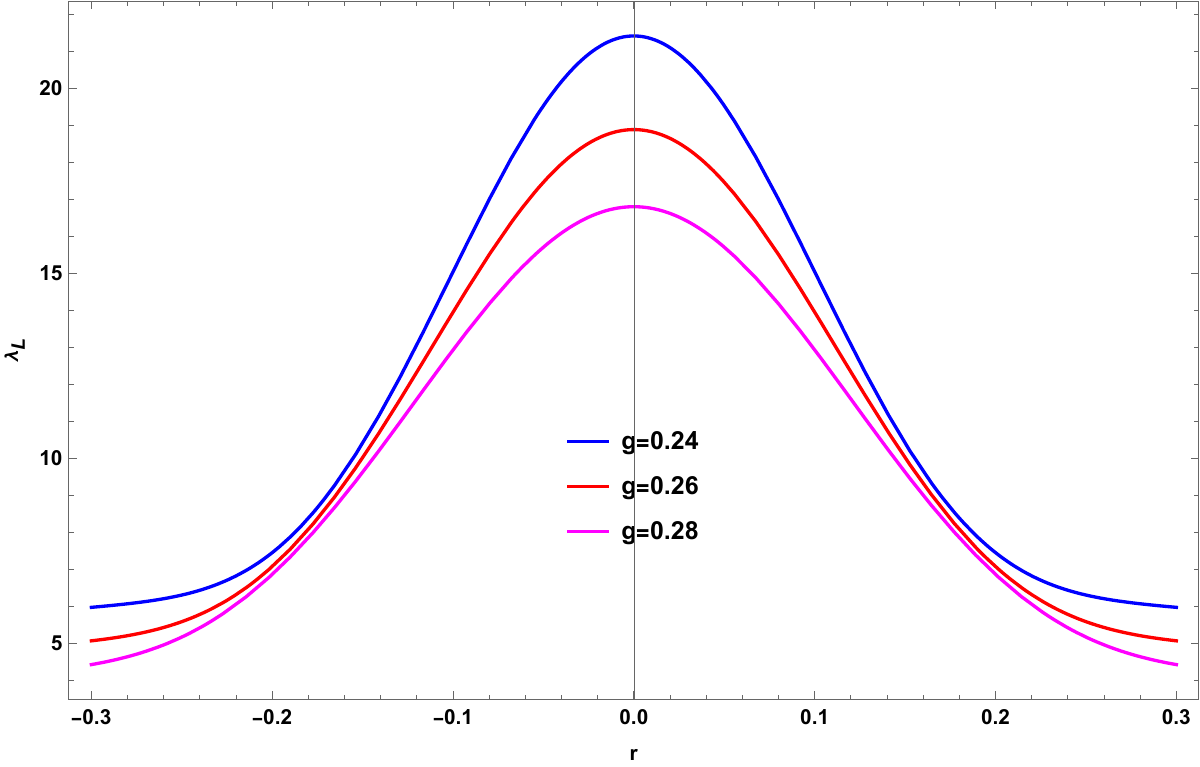}\\
    \includegraphics[width=0.45\linewidth]{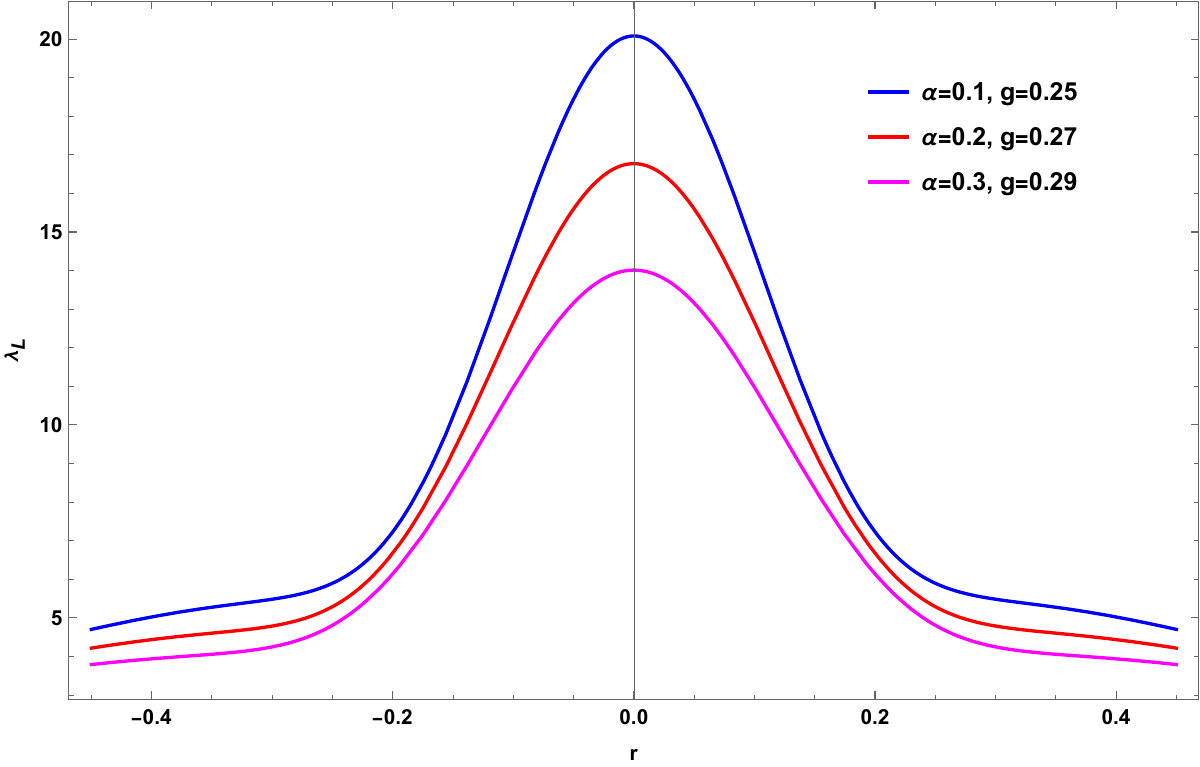}
    \caption{The radial dependence of the Lyapunov exponent $\lambda_{\mathrm{L}}$ as a function of $r$ for various values of the CS and the MM charge. Top left: $g=0.5$, Top right: $\alpha=0.1$. Here, we set $M=1$.}
    \label{fig:9c}
\end{figure}

In Figure \ref{fig:9c}, the radial dependence of the Lyapunov exponent for various values of the MM charge ($g$) and CS parameters ($\alpha$) is plotted. The curves represent different combinations of the MM charge and CS parameter. In the top-left panel, it is observed that as the cosmic string parameter $\alpha$ increases, the Lyapunov exponent $\lambda_{\mathrm{L}}$ decreases. Similarly, in the top-right panel, the Lyapunov exponent $\lambda_{\mathrm{L}}$ also decreases as the MM charge $g$ increases. Notably, when both the MM charge $g$ and the CS parameter $\alpha$ are increased simultaneously, the Lyapunov exponent $\lambda_{\mathrm{L}}$ continues to decrease with increasing values of both parameters. From this figure, it is clear that the time-like particle orbits are unstable or chaotic for the chosen values of both $g$ and $\alpha$. The positive Lyapunov exponent indicates the divergence between neighboring orbits, signaling the emergence of chaos in the system.

\section{\large \bf Scalar perturbations of ABG NLED-CS BH: The Klein-Gordon equation} \label{sec:3}

In this part, perturbations of zero-spin scalar field by deriving the Klein-Gordon equation around the selected BH solution is studied. The dynamics of massless scalar field, denoted as $\Phi$ are governed by the covariant form of the Klein-Gordon equation, which describes the evolution of a scalar field in a curved spacetime. This equation serves as the fundamental equation of motion for the scalar field in a gravitational background.

The Klein-Gordon equation for a massless scalar field in a general curved spacetime is given by the following wavew equation equation \cite{LL,RAK1,TR}:
\begin{equation}
    \frac{1}{\sqrt{-g}}\,\partial_{\mu}\,\left(\sqrt{-g}\,g^{\mu\nu}\,\partial_{\nu}\,\Phi\right)=0,\label{pp1}
\end{equation}
where $g_{\mu\nu}$ is the metric tensor, $g=\mbox{det}(g_{\mu\nu})$ is the determinant of the metric tensor, and $\partial_{\mu}$ is the partial derivative with respect to the coordinates system.

For the chosen BH spacetime (\ref{bb1}), we perform the following coordinate change (called tortoise coordinate) 
\begin{equation}
    dr_*=\frac{dr}{f(r)},\quad\quad \partial_{r_*}=f(r)\,\partial_r.\label{pp2}
\end{equation}
Thereby, the line-element Eq. (\ref{bb1}) becomes
\begin{equation}
    ds^2=f(r_*)\,\{-dt^2+dr^2_{*}\}+h^2(r_*)\,(d\theta^2+\sin^2 \theta\,d\phi^2),\label{pp3}
\end{equation}
where the metric tensor now
\begin{eqnarray}
    &&g_{\mu\nu}=\mbox{diag}\left(-f(r_*),\, f(r_*),\, h(r_*),\, h(r_*)\,\sin^2\theta\right),\label{pp4}\\
    &&g=\mbox{det}(g_{\mu\nu})=-f^2(r_*)\,h^4(r_*)\,\sin^2\theta.\label{pp5}
\end{eqnarray}

Moreover, we consider an ansatz for the scalar field $\Phi (t, r_*, \theta, \phi)$ of the following form: 
\begin{equation}
    \Phi(t, r_{*},\theta, \phi)=\exp(i\,\omega\,t)\,Y^{m}_{\ell} (\theta,\phi)\,\frac{\psi(r_*)}{r_*},\label{pp6}
\end{equation}
where $Y^{\ell}_{m} (\theta,\phi)$ are the spherical harmonics, $\omega$ is (possibly complex) temporal frequency in the Fourier domain \cite{TR}, and $\psi (r_*)$ is a propagating scalar field in the candidate spacetime.

Explicitly writing Eq. (\ref{pp1}) using (\ref{pp5}) and (\ref{pp6}), we find the Schrodinger-like wave equation:
\begin{equation}
    \frac{\partial^2 \psi(r_*)}{\partial r^2_{*}}+\left\{\omega^2-\mathcal{V}_0\right\}\,\psi(r_*)=0,\label{pp7}
\end{equation}
where $\mathcal{V}_0$ is the scalar perturbative potential given by
\begin{eqnarray}
    \mathcal{V}_0&=&f(r)\,\left(\frac{\ell\,(\ell+1)}{r^2}+\frac{f'(r)}{r}\right)\nonumber\\
    &=&\left(1-\alpha-\frac{2\,M\,r^2}{\left(r^2+g^2\right)^{3 / 2}}+\frac{g^2\,r^2}{\left(r^2+g^2\right)^2}\right)\left(\frac{\ell\,(\ell+1)}{r^2}+\frac{2\,M\,(r^2-2\,g^2)}{(r^2+g^2)^{5/2}}-\frac{2\,g^2\,(r^2-g^2)}{(r^2+g^2)^3}\right).\label{pp8}
\end{eqnarray}

The expression for the scalar perturbation potential indicates that the presence of the MM charge parameter, denoted by $g$, and the CS parameter $\alpha$ has a significant impact on the structure of this potential. These parameters introduce a distortion in the spacetime geometry, which in turn modifies the nature of the potential. Furthermore, the multipole number $\ell$, contributes additional modifications. The multipole number reflects the influence of higher-order angular momentum components, which further alter the potential’s form. As a result, the combined effects of $g, \alpha$ and $\ell$ lead to shifts in the potential for scalar fields having zero-spin value.

\begin{figure}[htbp]
    \centering
    \includegraphics[width=0.4\linewidth]{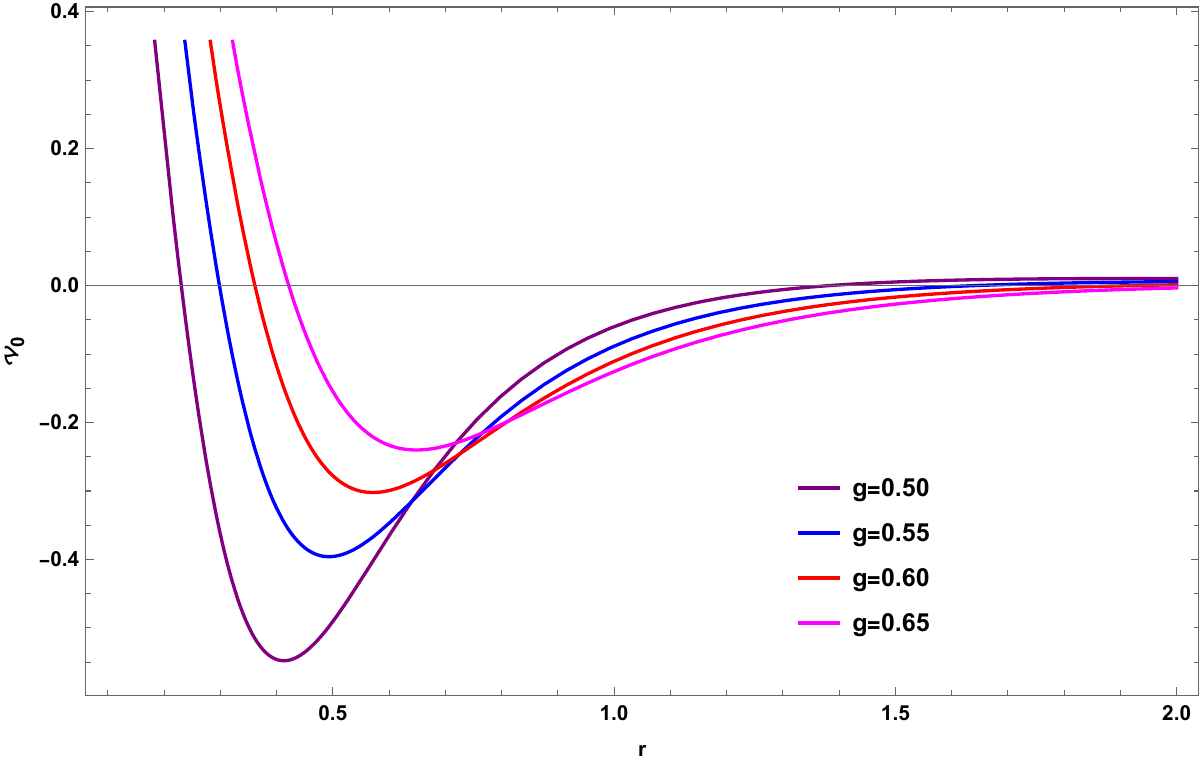}\quad\quad
    \includegraphics[width=0.4\linewidth]{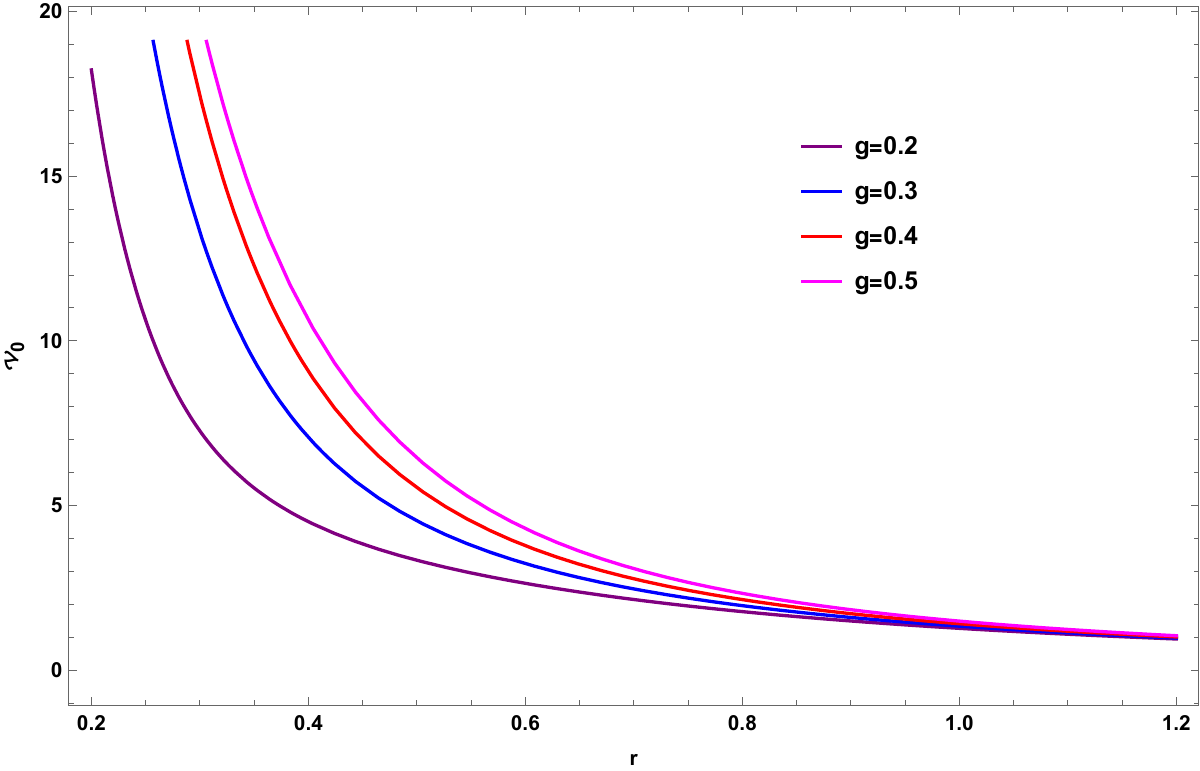}
    \caption{The behavior of scalar perturbation potential $\mathcal{V}_0$ as a function of $r$ for various values of $g$. Here, $M=0.2, \alpha=0.1$. Left panel: $\ell=0$, Right panel: $\ell=1$.}
    \label{fig:9}
    \centering
    \includegraphics[width=0.4\linewidth]{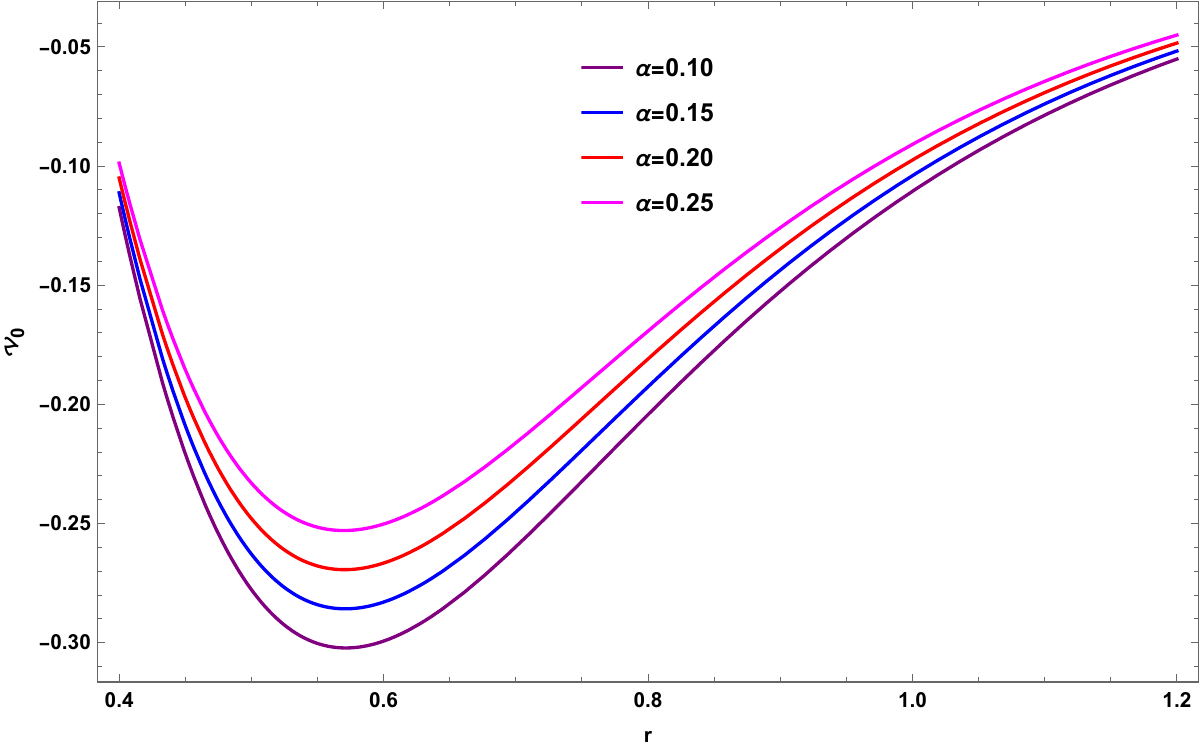}\quad\quad
    \includegraphics[width=0.4\linewidth]{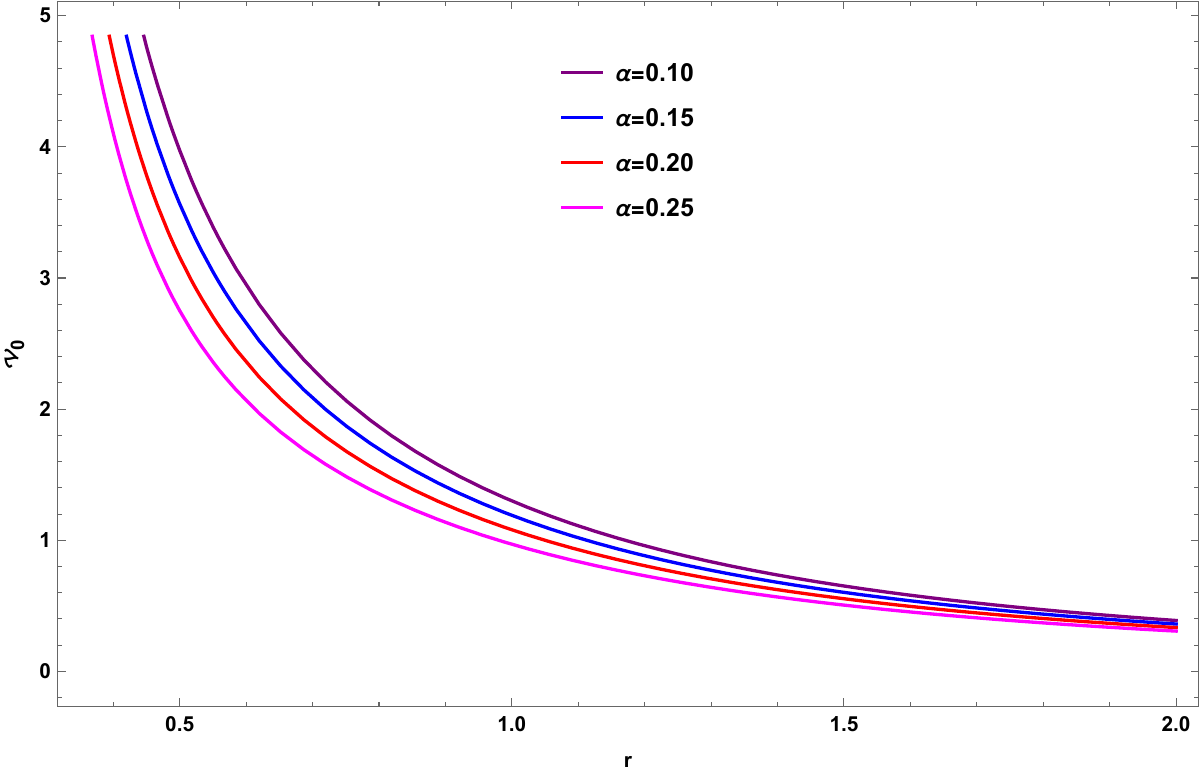}
    \caption{The behavior of scalar perturbation potential $\mathcal{V}_0$ as a function of $r$ for various values of $\alpha$. Here, $M=0.2$. Left panel: $\ell=0, g=0.6$, Right panel: $\ell=1, g=0.25$.}
    \label{fig:10}
    \centering
    \includegraphics[width=0.4\linewidth]{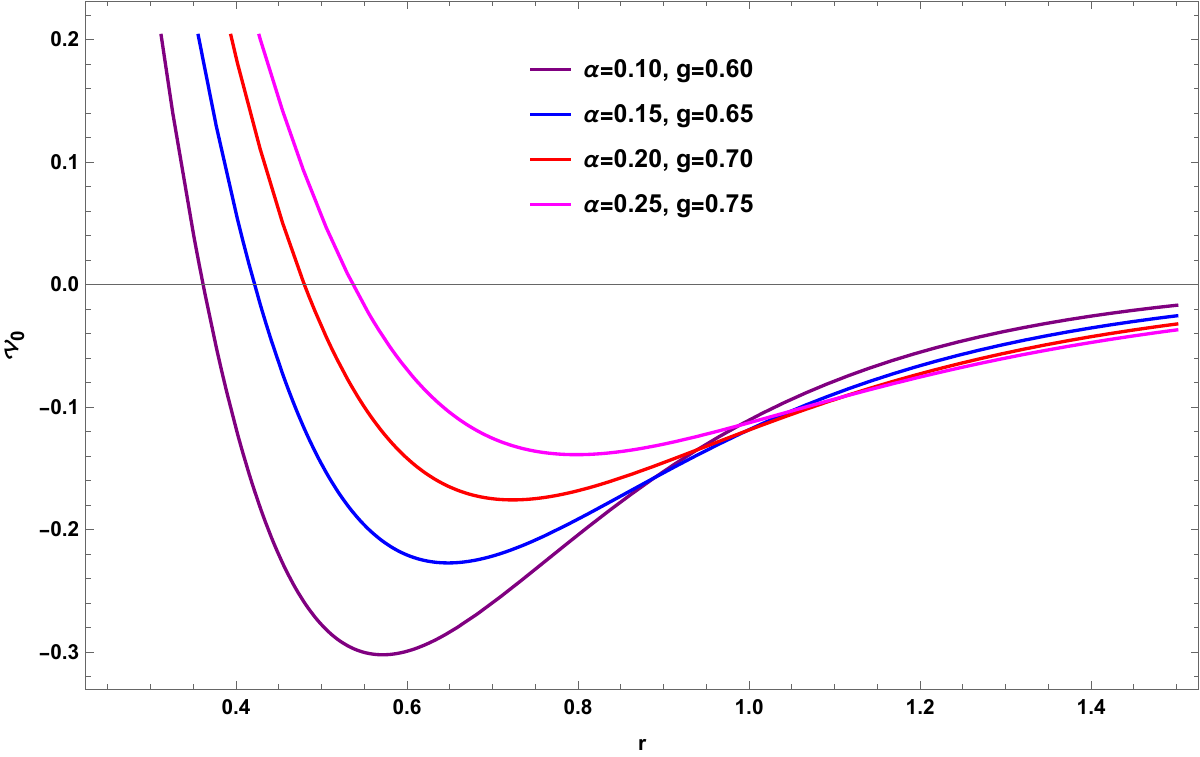}\quad\quad
    \includegraphics[width=0.4\linewidth]{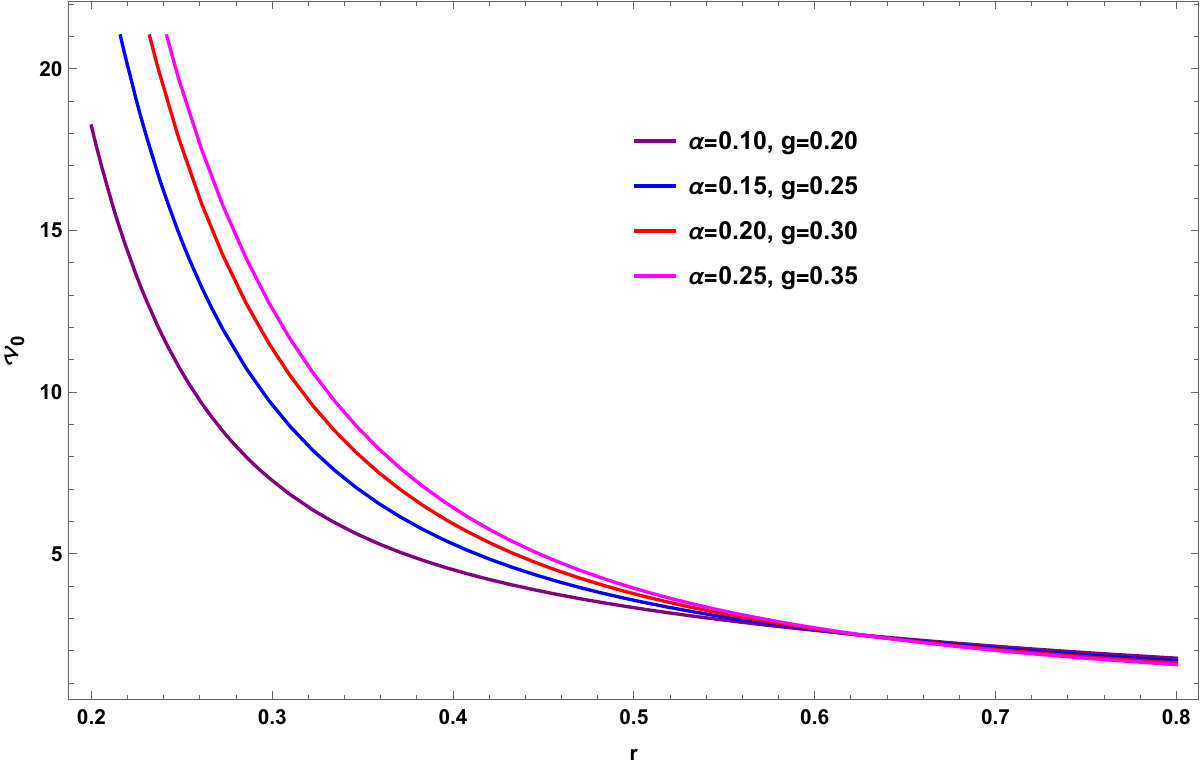}
    \caption{The behavior of scalar perturbation potential $\mathcal{V}_0$ as a function of $r$ for various values of $\alpha, g$. Here, $M=0.2$. Left panel: $\ell=0$, Right panel: $\ell=1$.}
    \label{fig:11}
\end{figure}

We have generated Figures \ref{fig:9}--(\ref{fig:11}), which illustrates the behavior of the scalar potential (\ref{pp8}) for spin-zero scalar fields, varying the MM charge $g$ and CS parameter $\alpha$ for multipole numbers $\ell=0$ and $\ell=1$. These figures clearly show that for the multipole number $\ell=0$, increasing the values of both $g$ and $\alpha$ leads to an increase in the scalar potential. In contrast, for the multipole $\ell=1$, we observe a decreasing trend in the scalar potential. This demonstrates how the inclusion of higher multipole numbers reduces the potential, emphasizing the significant role that the multipole number plays in shaping the profile of the scalar potential in the presence of ABG NLED-CS BH.

To examine the qualitative features of the scalar potential, it is convenient to reexpress this potential in terms of dimensionless variables. Defining the parameters $x=r/M$ and $y=g/M$, we find the following dimensionless quantity from Eq. (\ref{pp8}) as,
\begin{eqnarray}
    M^2\,\mathcal{V}_0=\left(1-\alpha-\frac{2\,x^2}{\left(x^2+y^2\right)^{3 / 2}}+\frac{x^2\,y^2}{\left(x^2+y^2\right)^2}\right)\left(\frac{\ell\,(\ell+1)}{x^2}+\frac{2\,(x^2-2\,y^2)}{(x^2+y^2)^{5/2}}-\frac{2\,y^2\,(x^2-y^2)}{(x^2+y^2)^3}\right).\label{pp9}
\end{eqnarray}
For a scalar s-wave where the multipole number $\ell=0$, we find from Eq. (\ref{pp9})
\begin{eqnarray}
    M^2\,\mathcal{V}_0\Big{|}_{\ell=0}=\left(1-\alpha-\frac{2\,x^2}{\left(x^2+y^2\right)^{3 / 2}}+\frac{x^2\,y^2}{\left(x^2+y^2\right)^2}\right)\left(\frac{2\,(x^2-2\,y^2)}{(x^2+y^2)^{5/2}}-\frac{2\,y^2\,(x^2-y^2)}{(x^2+y^2)^3}\right).\label{pp10}
\end{eqnarray}

\begin{figure}[htbp]
    \centering
    \includegraphics[width=0.45\linewidth]{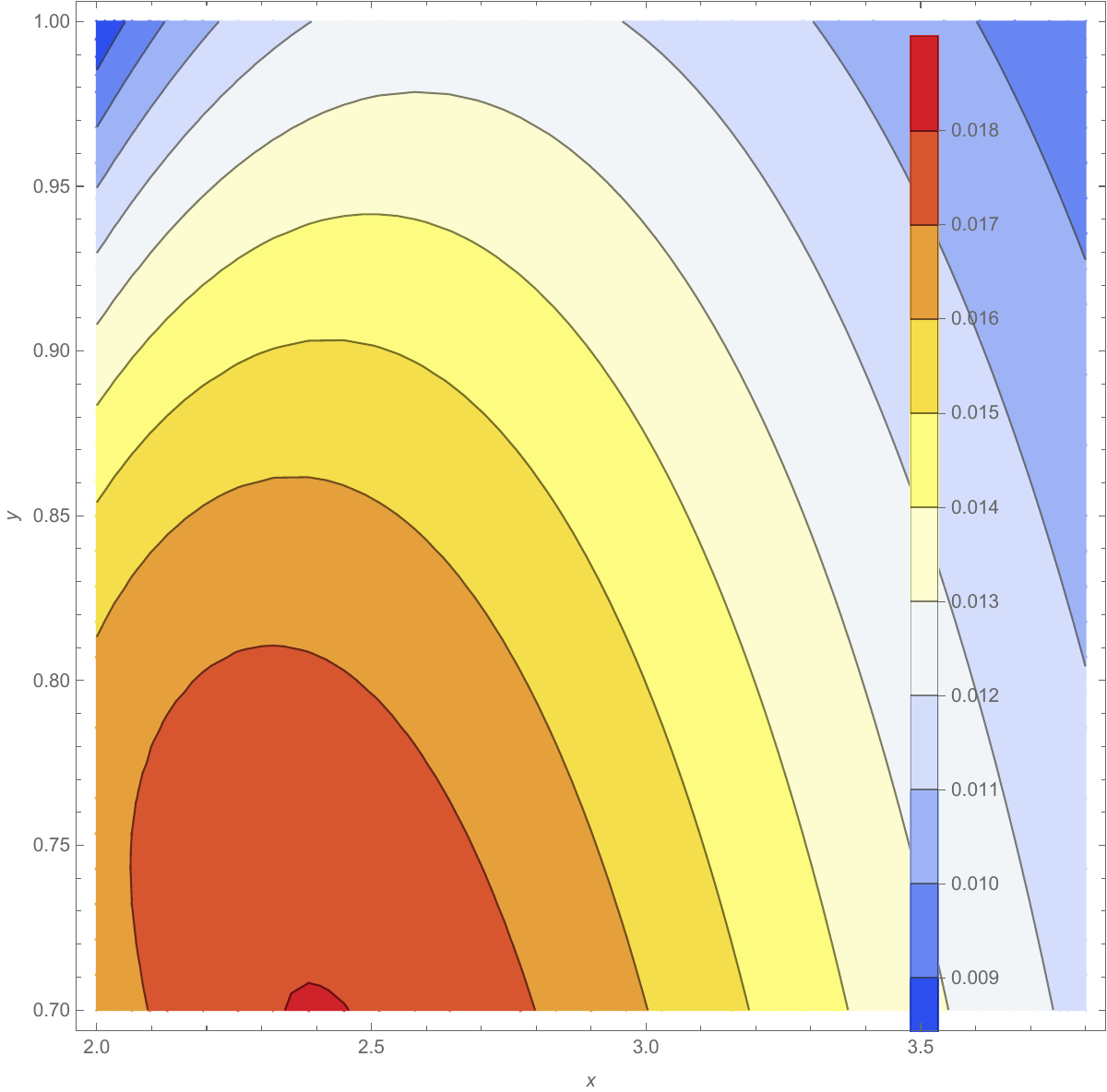}
    \caption{The qualitative feature of zero spin scalar potential for the dominant multipole number $\ell=0$ and $\alpha=0.1$ is depicted in Contour plot. 'Red' to 'blue' corresponds to 'high' to 'low'.}
    \label{fig:14}
\end{figure}

The most notable feature of zero spin is the peak associated with the spin-zero state. By drawing a line that connects the peaks in Figure \ref{fig:14}, we can represent a 'line of best fit.' From this figure, we can approximate $r \approx \bar{r}$ which can be used to determine the relevant QNM profiles. In the current study we discard this since QNMS already discussed in Ref. \cite{AK1}.

\begin{figure}[htbp]
    \includegraphics[width=0.40\linewidth]{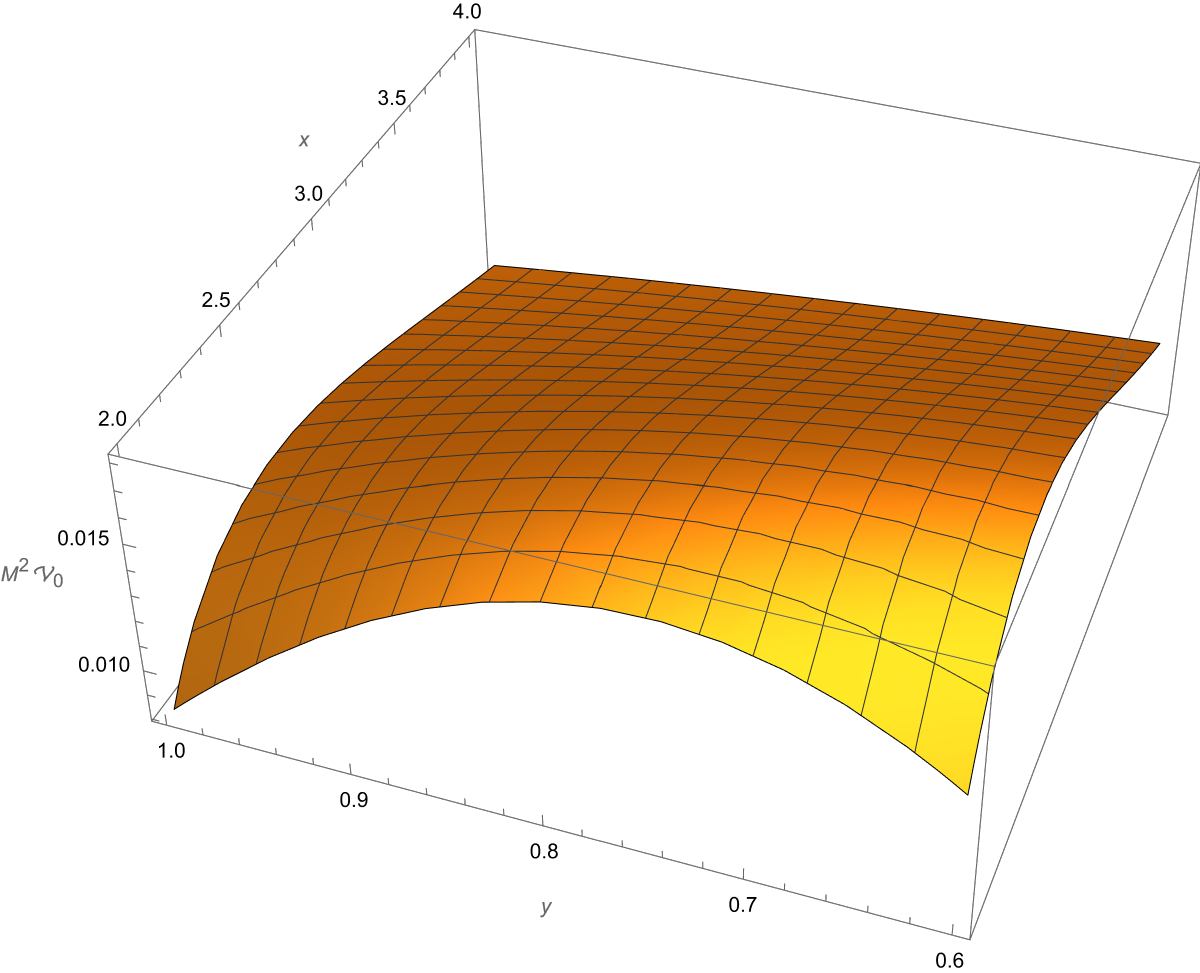}\quad\quad
    \includegraphics[width=0.53\linewidth]{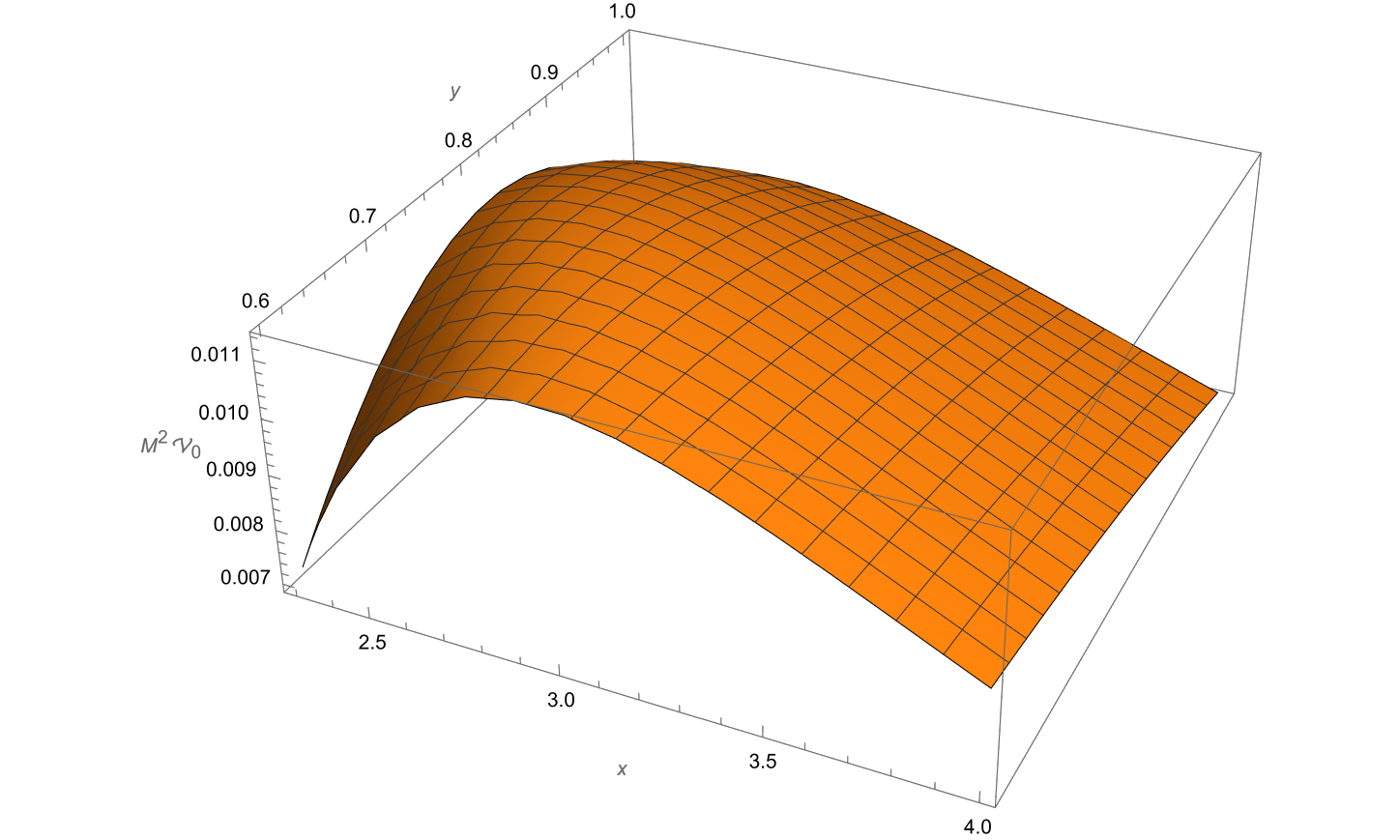}
    \caption{Three-dimensional plot of $M^2\,\mathcal{V}_0$: the qualitative features of zero spin scalar potential for the dominant multipole number $\ell=0$. Left panel: $\alpha=0.1$, right panel: $\alpha=0.2$}
    \label{fig:12}
\end{figure}

\begin{figure}[htbp]
    \includegraphics[width=0.5\linewidth]{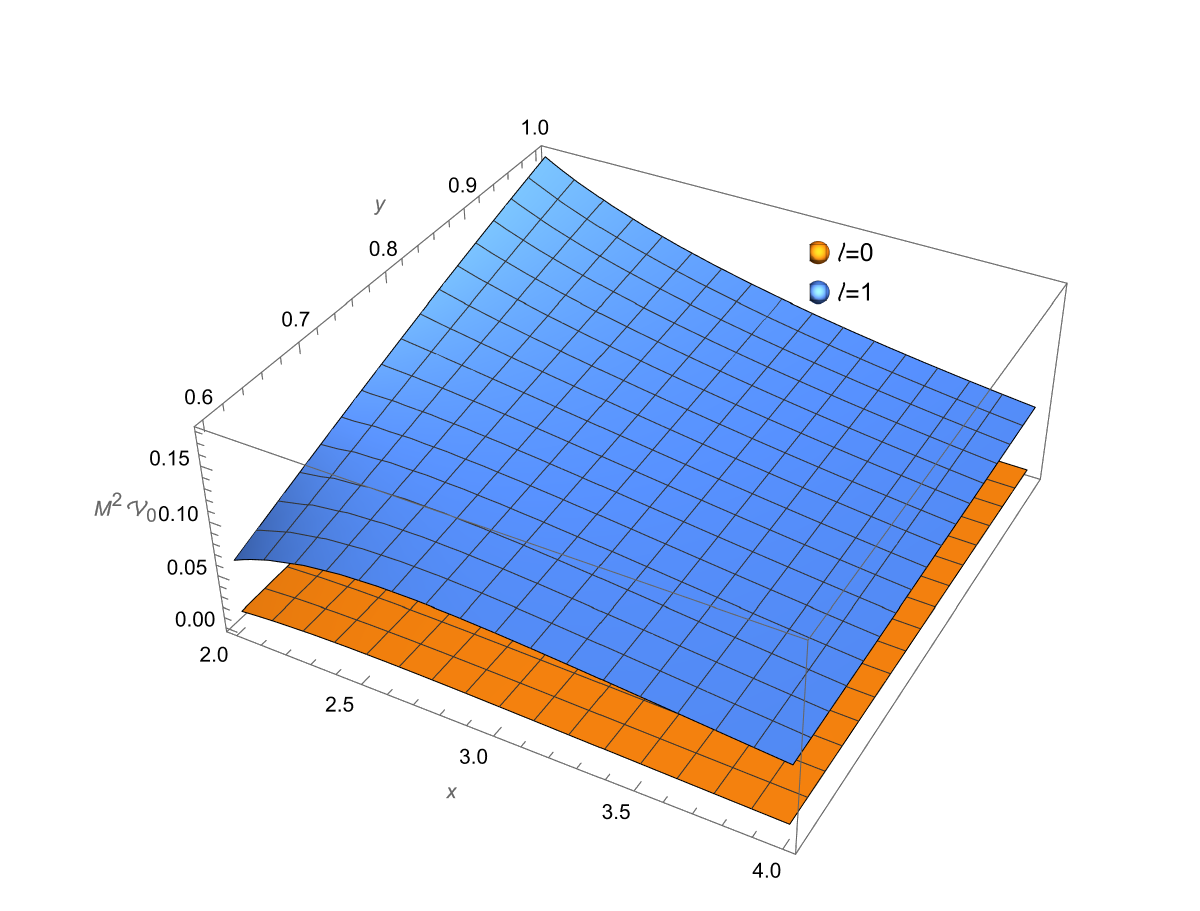}\quad\quad
    \includegraphics[width=0.5\linewidth]{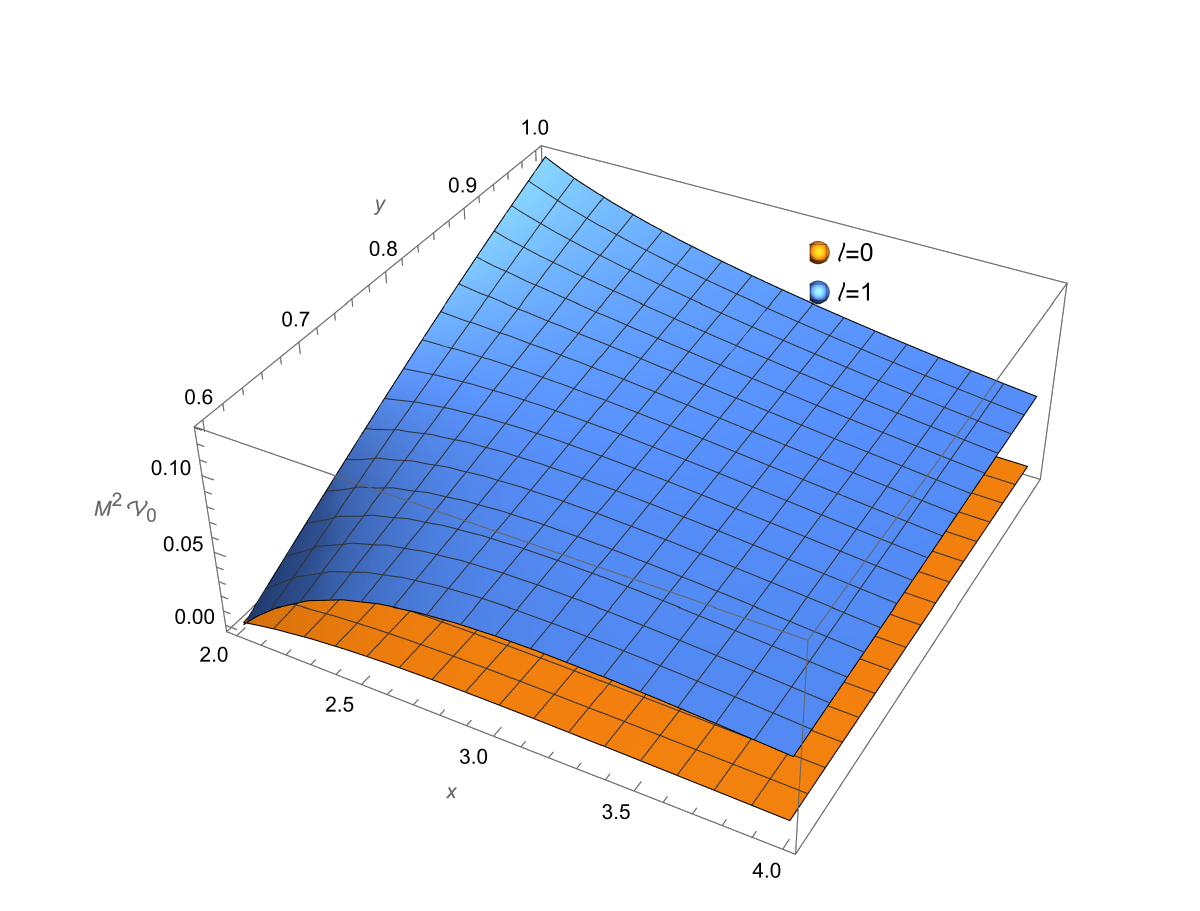}
    \caption{Comparison of three-dimensional plot of $M^2\,\mathcal{V}_0$ for the multipole number $\ell=0$ and and $\ell=1$: the qualitative features of zero spin scalar potential. Left panel: $\alpha=0.1$, right panel: $\alpha=0.2$}
    \label{fig:13}
\end{figure}

In Figure \ref{fig:12}, we illustrate a three-dimensional graph of the dimensionless quantity $M^2\,\mathcal{V}_0$, dimensionless parameters $x$ and $y$ for different values of the CS parameter $\alpha=0.1$ (left panel) and $\alpha=0.2$ (right panel) in the dominant mode $\ell=0$. Similarly, in Figure \ref{fig:13}, similar illustration of the dimensionless quantity $M^2\,\mathcal{V}_0$, dimensionless parameters $x$ and $y$ for the multipole number $\ell=0$ and $\ell=1$ with the CS parameter $\alpha=0.1$ (left panel) and $\alpha=0.2$ (right panel).

\subsection{\large \bf Greybody Factors (GFs) of BH} \label{sec:3.1}

The purpose of this section is to compute the transmission coefficient, or the so-called GF, for the ABG NLED-CS BH space-time, taking into account the effects of MM charge $g$. The GF is a value that indicates how far the radiation spectrum deviates from black body radiation. Researches on reflection and transmission coefficients have been conducted using several methodologies \cite{gf1,gf2,gf3,gf4,gf5,gf6}. To examine the GFs, we shall use the general semi-analytic bounds method. This method requires GFs to be greater than or equal to the formula \cite{gf3,gf7,gf8}. 
\begin{equation}
{T}\left( \omega\right) \geq \sec h^{2}\left( \int_{-\infty }^{+\infty }\wp
dr_{\ast }\right) ,  \label{is8}
\end{equation}%
where ${T}\left( \omega\right) $ is the transmission probability or GF, $r_{\ast }$
is the tortoise coordinate and $\wp $ is given by:
\begin{equation}
\wp =\frac{1}{2h}\sqrt{\left( \frac{dh\left( r_{\ast }\right) }{dr_{\ast }}%
\right) ^{2}+(\omega^{2}- \mathcal{V}-h^{2}\left( r_{\ast }\right) )^{2}}.  \label{is9}
\end{equation}
where $ \mathcal{V}$ is the potential  and $h(r_{\ast })$ is a positive function satisfying $h\left( -\infty
\right) =h\left( \infty \right) = \omega$. We
select $h= \omega $. Therefore, Eq. (\ref{is8}) becomes:
\begin{equation}
   T({w}) \geq \sec h^2\left(\frac{1}{2\,{w}}\int^{+\infty}_{-\infty}\,\mathcal{V}_0\,dr_{*}\right).\label{qq1}  
\end{equation}

\begin{figure}[htbp]
    \centering
    \includegraphics[width=0.43\linewidth]{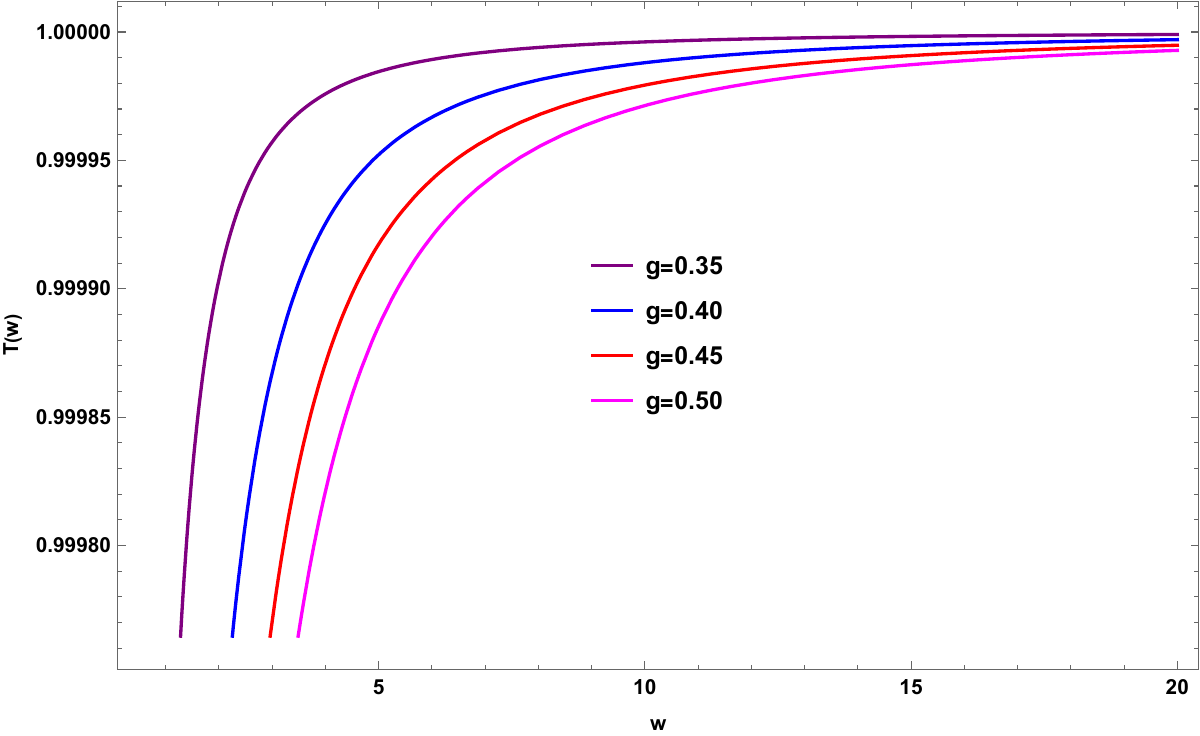}\quad\quad
    \includegraphics[width=0.41\linewidth]{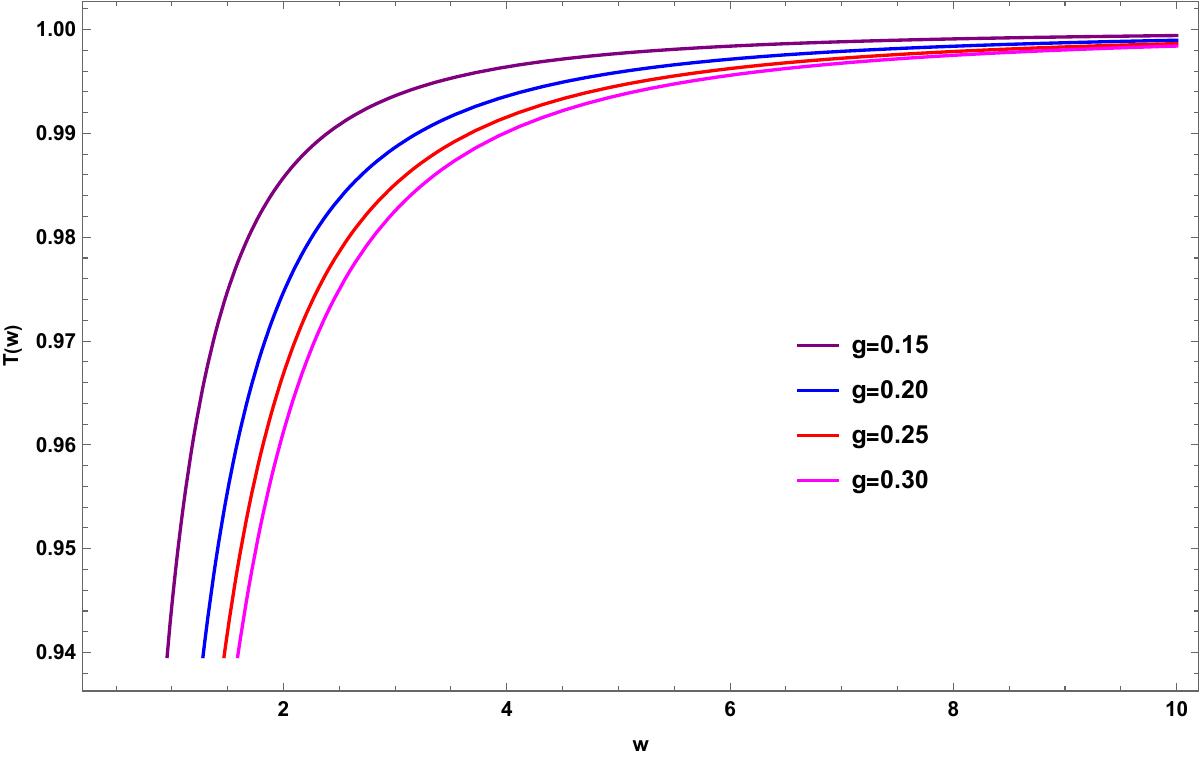}
    \caption{The GF $T$ as a function of $w$ for different values of MM charge parameter $g$. Here, we set $M=1$ and a numerical value of $r_\text{h}=2.5\,M$. Left panel: $\ell=0$, right panel: $\ell=1$.}
    \label{fig:5a}
\end{figure}

\begin{figure}[htbp]
    \centering
    \includegraphics[width=0.45\linewidth]{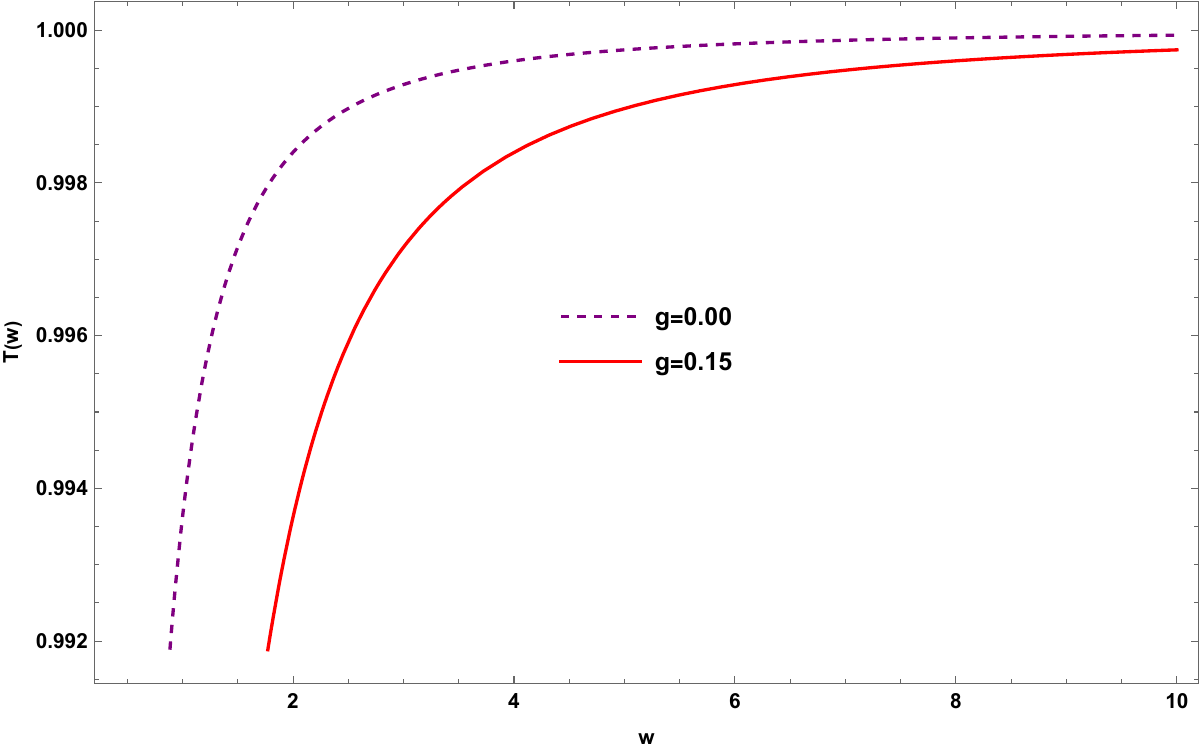}
    \caption{Comparison of the GF $T(w)$ as a function of $w$. Here dashed line correspond to zero MM charge, $g=0$, and hence horizon radius $r_\text{h}=\frac{2\,M}{1-\alpha}=2.5\,M$ with $\alpha=0.2$ and colored line correspond to the selected BH solution where we choose a numerical value of $r_\text{h}=2.5\,M$. Here, we set $M=1$ and $\ell=0$.}
    \label{fig:5b}
\end{figure}

Using Eqs. (\ref{pp2}) and (\ref{pp8}), Eq. (\ref{qq1}) becomes
\begin{equation}
   T({w}) \geq \sec h^2\left[\frac{1}{2\,{w}}\int^{+\infty}_{r_\text{h}}\,\left(\frac{\ell\,(\ell+1)}{r^2}+\frac{2\,M\,(r^2-2\,g^2)}{(r^2+g^2)^{5/2}}-\frac{2\,g^2\,(r^2-g^2)}{(r^2+g^2)^3}\right)\,dr\right].\label{qq2}  
\end{equation}
Now, different integrals are as follows:
\begin{eqnarray}
    &&\int^{+\infty}_{r_\text{h}}\,\frac{\ell\,(\ell+1)}{r^2}\,dr=\frac{\ell\,(\ell+1)}{r_\text{h}},\nonumber\\
    &&\int^{+\infty}_{r_\text{h}}\,\frac{2\,M\,(r^2-2\,g^2)}{(r^2+g^2)^{5/2}}\,dr=\frac{2\,M}{g^2}\,\left(\frac{r^3_\text{h}+2\,r_\text{h}\,g^2}{(r^2_\text{h}+g^2)^{3/2}}-1\right),\nonumber\\
    &&\int^{+\infty}_{r_\text{h}}\,\frac{2\,g^2\,(r^2-2\,g^2)}{(r^2+g^2)^3}\,dr=\frac{1}{2\,g}\,\left(\frac{r_\text{h}\,(7\,g^2+r^2_\text{h})/2}{(r^2_\text{h}+g^2)^2}-\cot^{-1} \left(\frac{r_\text{h}}{g}\right)\right).\label{qq3}
\end{eqnarray}

It is important to note that the transmission coefficients, although not dependent on the CS parameter, are solely determined by the MM charge $g$. Furthermore, the numerical value of the radius of the horizon $r_\text{h}$ depends on the CS parameter. Figure \ref{fig:5a} shows the behavior of the greybody factor for a scalar perturbative potential of a spin-0 quantum field, focusing on how the transmission coefficients vary with changes in the MM charge $g$. As $g$ increases, the transmission coefficients decrease, indicating that less thermal radiation reaches the observer at spatial infinity. In the figure, we set the radius of the horizon to $r_\text{h}=2.5\,M$ and consider two different multipole numbers: $\ell = 0$ (left panel) and $\ell = 1$ (right panel). Figure \ref{fig:5b} presents a comparison of GF with and without MM charge $g$ for multipole number $\ell = 0$. It should be noted that for $g=0$ (that is, no MM charge), the horizon radius is $r_\text{h} = \frac{2\,M}{1-\alpha} = 2.5\,M$, where we have chosen $\alpha = 0.2$.

\section{\large Summary and Conclusions}\label{sec:4}

In this work, we analyzed the geodesic motion and scalar perturbations of the ABG NLED-CS BH. The study primarily focused on the influence of the MM charge $g$ and the CS parameter $\alpha$ on the effective potential, photon trajectories, dynamics of massive and massless test particles, and scalar perturbations.
We derived the effective potential for null and time-like geodesics, demonstrating how $g$ and $\alpha$ alter the potential compared to standard BH solutions. The results highlighted that increasing $g$ and $\alpha$ leads to significant changes in the effective potential, as illustrated in Figures \ref{fig:1} and \ref{fig:2}. Specifically, for null geodesics, the potential increased with higher values of $g$ and $\alpha$, thereby modifying the trajectories of massless particles. Similarly, time-like geodesics showed an increase in the effective potential under the influence of these parameters. These modifications were crucial in understanding the gravitational behavior near the ABG NLED-CS BH. The photon trajectory equation was derived and analyzed, revealing that $g$ and $\alpha$ collectively modify the geodesic paths of photons. The results emphasized that these parameters introduce deviations from the standard Schwarzschild BH solution. Figures \ref{fig:3} and \ref{figa9} provided a comprehensive visual representation of photon and massive particle trajectories, showing precession and orbital shifts due to the CS parameter and MM charge. Notably, the angular momentum and energy of massive test particles in circular orbits were significantly influenced by $g$ and $\alpha$, as demonstrated in Figures \ref{fig:6}--\ref{fig:8}. The ISCO radius was shown to increase with $\alpha$ and decrease with $g$, highlighting the opposing effects of these parameters, as tabulated in Table \ref{taba1} and shown in Figure \ref{figa1}. Furthermore, the ISCO analysis demonstrated that the MM charge $g$ and the CS parameter $\alpha$ had opposing effects on the orbital stability of massive test particles. While $g$ reduced the ISCO radius, $\alpha$ caused an increase, leading to a complex interplay between these parameters. The results tabulated in Table \ref{taba1} and depicted in Figure \ref{figa1} provided a quantitative understanding of these effects. Moreover, the angular momentum and energy of particles in circular orbits, as shown in Figures \ref{fig:6}--\ref{fig:8}, revealed significant modifications due to $g$ and $\alpha$. These findings demonstrated how the presence of NLED and CS contributions alters the classical dynamics of test particles. Moreover, to assess the stability of time-like particle orbits, we generated Figure \ref{fig:9c}, which illustrates the Lyapunov exponent $\lambda_{\mathrm{L}}$ as a function of the radial coordinate $r$ for different values of the MM charge $g$ and the CS parameter $\alpha$. The results clearly demonstrate that the time-like particle orbits are unstable for the selected ABG NLED-CS BH solution.

In the analysis of GFs, we employed a semi-analytic bounds method to compute the transmission coefficients for the ABG NLED-CS BH. The results, shown in Figures \ref{fig:5a} and \ref{fig:5b}, indicated that the MM charge $g$ reduced the transmission coefficients, thereby lowering the probability of scalar radiation escaping to spatial infinity. This behavior underscored the role of the MM charge in modifying the thermal radiation properties of the BH. The numerical computations further confirmed the impact of $g$ and $\alpha$ on the GF, with significant deviations observed from the standard BH cases. We also studied scalar perturbations around the ABG NLED-CS BH by deriving the Klein-Gordon equation for a massless scalar field. The scalar perturbative potential was found to depend on $g$, $\alpha$, and the multipole number $\ell$. Figures \ref{fig:9}--\ref{fig:11} illustrated the variation of the scalar potential with these parameters, revealing that higher values of $g$ and $\alpha$ increased the potential for $\ell=0$, while $\ell=1$ exhibited a decreasing trend. Additionally, three-dimensional plots (Figures \ref{fig:12} and \ref{fig:13}) provided further insights into the qualitative features of the scalar potential in dimensionless variables. These results highlighted the intricate interplay between NLED and the CS parameter in shaping the dynamics of scalar fields. The contour and three-dimensional plots (Figures \ref{fig:12} and \ref{fig:13}) further illustrated the qualitative features of the scalar potential, emphasizing the significant role of NLED and CS in shaping the perturbative dynamics

In conclusion, this study highlighted the significant impact of the MM charge and CS parameter on the geodesic motion, scalar perturbations, and greybody factors of the ABG NLED-CS BH. These findings deepen our understanding of the interplay between NLED and cosmic strings in BH physics. Future work could explore the stability of these solutions under higher-order perturbations, investigate their observational signatures in astrophysical settings, and analyze their thermodynamic properties in greater detail, particularly in the context of quantum corrections and alternative gravity theories. Additionally, the influence of higher multipole moments and the inclusion of spin effects could provide further insights into the complex dynamics of this system.

\section*{Data Availability Statement}

No new data were generated or analyzed in this study.

\section*{Conflicts of Interest}

There is no conflict of interests.



\section*{Code/Software}

No software/Coder were used in this study.

\section*{Acknowledgments}

F.A. acknowledges the Inter University Centre for Astronomy and Astrophysics (IUCAA), Pune, India for granting visiting associateship. \.{I}. S. expresses gratitude for the networking assistance provided by COST Actions CA21106, CA22113, and CA23130, as well as to T\"{U}B\.{I}TAK, ANKOS, and SCOAP3 for their support.

\end{document}